\documentclass[
reprint,
 amsmath,amssymb,
 aps,
prb,
]{revtex4-2}

\usepackage{adjustbox}
\usepackage[normalem]{ulem}
\usepackage{float}
\usepackage{amssymb,amsmath,amsfonts,tabularx}
\usepackage{physics}
\usepackage{tikz}
\usetikzlibrary{calc}
\usetikzlibrary{matrix}
\usepackage[page]{appendix}
\usepackage{blindtext}
\usepackage{titlesec}
\usepackage{color}
\usepackage{soul}
\usepackage{dcolumn}
\usepackage{bm}
\usepackage{chngcntr}
\usepackage{multirow}
\usepackage{comment}

\begin{document}

\preprint{APS/123-QED}

\title{Quantum skyrmions in the antiferromagnetic triangular lattice}


\author{In\'es Corte$^{1,2}$}
\email{corresponding author: ines.corte@iflp.unlp.edu.ar}
\author{Federico Holik$^3$}
\author{Lorena Reb\'on$^{1,2}$}
\author{Flavia A. G\'omez Albarrac\'in$^{4,2}$}
\affiliation{
 $^1$Instituto de F\'isica La Plata (CONICET-UNLP). Diagonal 113 S/N, La Plata, Argentina
}
\affiliation{
 $^2$Departamento de Ciencias B\'asicas, Facultad de Ingenier\'ia, Universidad Nacional de La Plata (UNLP). Avenida 1 y 47 S/N, La Plata, Argentina
}
\affiliation{
 $^3$Universidad Nacional de Hurlingham (UNAHUR), Laboratorio de Investigación y Desarrollo Experimental en Computación (LIDEC), Hurlingham, Argentina
}
\affiliation{
 $^4$Instituto de F\'isica de L\'iquidos y Sistemas Biol\'ogicos (CONICET-UNLP). Calle 59 789, La Plata, Argentina
}



\date{\today}

\begin{abstract}
Magnetic skyrmions are topological quasiparticles potentially useful for memory and computing devices. Antiferromagnetic (AF) skyrmions present no transverse deflection, making them suitable candidates for data storage applications.
After the discovery of skyrmions with length scales comparable to the lattice constant, several works presented quantum analogues of classical ferromagnetic skyrmions in spin systems. However, studies about quantum analogues of AF skyrmions are still lacking.
Here, we explore the phases of the AF quantum spin-1/2 Heisenberg model with Dzyaloshinskii-Moriya interactions on the triangular lattice using the density matrix renormalization group (DMRG) algorithm. We study the magnetization profile, spin structure factor and quantum entanglement of the resulting ground states to characterize the corresponding phases and signal the emergence of quantum AF skyrmions. {Moreover, we present calculations of the quantum analogs of the scalar chirality, and construct local chirality maps, to further explore the different phases.} Our results support that three-sublattice quantum antiferromagnetic skyrmion textures are stabilized in a wide range of magnetic fields. 
\end{abstract}

\maketitle


\section{\label{sec:intro} Introduction}
Magnetic skyrmions are swirling magnetic textures \cite{Bogdanov1989,Bogdanov1994} {with non trivial topology. The associated topological invariant in real space in the continuous picture is the topological charge, defined as} 

\begin{equation} \label{eq:tc}
{Q}=\frac{1}{4\pi} \int d^2x \: \mathbf{m}(\mathbf{r}) \cdot 
\left(\partial_x  \mathbf{m}(\mathbf{r}) \times \partial_y  \mathbf{m}(\mathbf{r}) \right),
\end{equation}
\noindent where $\mathbf{m}(\mathbf{r})$ is a local unit vector field associated to the spin vector at position $\mathbf{r}$. The topological charge {$Q$} counts how many times the texture wraps around a sphere, and it yields {$Q=-1$} for {core-down} skyrmions \cite{Tokura2021review}.

Due to their non-trivial topology, skyrmions are robust spin textures  and have therefore attracted considerable attention, particularly in connection with potential technological applications such as memory devices \cite{NagaosaTech, FertTech,SampaioTech}. These textures may be physically stabilized via different mechanisms \cite{beyond_skys}, ranging from magnetic frustation \cite{Okubo2012,Leonov2015,Amoroso2020} to long-range interactions \cite{Utesov2021}. One of the most relevant mechanisms to realize skyrmions in non-centrosymmetric systems is the in-plane Dzyaloshinskii-Moriya interaction (DMI), a key ingredient in the models connected to the first experimental measurements \cite{MnSi,Yu2010}. 

Skyrmions usually encountered in experimental measurements behave like classical particles. However, their size can range from micrometers to lengths comparable to the interatomic lattice spacing. In fact, atomic-scale skyrmions have been observed experimentally in several magnetic systems over the last years \cite{Heinze2011, Linder2020, Kono2026}. Since quantum effects can play an important role in the latter case, skyrmions with diameters of a few nanometers may not be well-described using classical spins. Several numerical studies of quantum analogs of ferromagnetic (F) skyrmions have been made over recent years \cite{SkyF_SW, Lohani2019, sky_ED, PhysRevB.108.094410, sky_NNs, PhysRevResearch.4.023111, Qsky}. These works state the emergence of quantum ferromagnetic skyrmions as ground states of model Hamiltonians using spin waves \cite{SkyF_SW}, exact diagonalization \cite{Lohani2019, sky_ED, PhysRevB.108.094410}, neural network quantum states \cite{sky_NNs, PhysRevResearch.4.023111} and density matrix renormalization group (DMRG) methods \cite{Qsky}. Most importantly, different measures of quantum entanglement between spins show that these ferromagnetic skyrmion states present entanglement \cite{Qsky, sky_NNs, PhysRevB.108.094410} and hence are indeed truly quantum. 
In this work, we focus instead on quantum properties of antiferromagnetic (AF) skyrmions.

One of the possible issues that may arise when considering technological applications is the so-called skyrmion Hall effect, which is the deviation of skyrmions when driven by a current. Different topological quasiparticles have been proposed as suitable candidates where this effect may be suppressed. Among them, of particular interest are the antiferromagntic skyrmions, which have been recently observed in a number of materials \cite{Gao2020,Rosales2022,AFSky1,AFSky2,SkyAFBubbles}
Quantum effects in these textures have been studied using techniques such as spin waves  \cite{spinwaves} and a modified quantum Monte Carlo method \cite{Liu_2020}. In this work, we explore the magnetic phases of ground states for a simple skyrmion Hamiltonian in the antiferromagnetic triangular lattice with $S=1/2$ spins obtained using the DMRG algorithm \cite{White92, White93, Schollwoek2011}. Although originally developed for one-dimensional quantum systems, this method has proven to be a powerful tool to compute ground states of two-dimensional lattices \cite{White96, hallberg, miles, rydbergkagome, skyliquid}. In particular, DMRG was recently used to successfully  simulate ground states hosting quantum ferromagnetic skyrmions \cite{Qsky, skyliquid}. Here, we follow a similar approach, showing evidence of the formation of skyrmions in the AF triangular lattice upon an external magnetic field, and studying quantum effects such as the entanglement.

 This article is organized as follows. In Section \ref{sec:model}, we specify the model Hamiltonian used in the DMRG calculations. This Hamiltonian gives rise to AF skyrmions in a classical-spins scenario. Our numerical findings are discussed in Section \ref{sec:results}, where we argue the emergence of quantum AF skyrmions as our main result. We calculate the magnetization and magnetic susceptibility (Section \ref{sec:magn}), {compute the structure factor (Section \ref{sec:Sq}) and analyze the total and local scalar chirality (Section \ref{sec:chirality})} of the ground states found with DMRG. {Additionally, we employ entanglement measures to characterize the AF skyrmion phase in Section \ref{sec:entanglement}. We also inspect whether the AF skyrmion texture is influenced by the boundary shape and lattice size in Section \ref{sec: stability}, and analize the robustness of the phase calculating bulk variables}. To conclude, we summarize our results in Section \ref{sec:conclusions}.

\section{Model and methods}
\label{sec:model}

The main goal of this work is to study the possible stabilization of skyrmions for $S=1/2$ spins in an antiferromagnetic model. To this end, we study the two-dimensional spin-1/2 Heisenberg model with DMI and antiferromagnetic exchange couplings $J>0$ in the triangular lattice. We employ the Hamiltonian

\begin{equation} \label{eq:H}
\hat{H}= \sum_{\langle\mathbf{r},\mathbf{r'}\rangle}[J\hat{\mathbf{S}}_{\mathbf{r}}\cdot \hat{\mathbf{S}}_{\mathbf{r'}} +\mathbf{D}_{\mathbf{r'},\mathbf{r}}\cdot (\hat{\mathbf{S}}_{\mathbf{r}}\times \hat{\mathbf{S}}_{\mathbf{r'}})]-\sum_{\mathbf{r}} B\hat{\mathbf{z}}\cdot \hat{\mathbf{S}}_{\mathbf{r}},
\end{equation}

\noindent where $\mathbf{\hat{S}}_\mathbf{r}=(\hat{S}^x_\mathbf{r}, \hat{S}_\mathbf{r} ^y, \hat{S}^z_\mathbf{r})$ ($\hat{S}^\alpha_\mathbf{r}$ with $\alpha=x,y,z$ are spin operators acting at position $\mathbf{r}$), $\mathbf{D}_{\mathbf{r'},\mathbf{r}}=D\hat{\mathbf{z}} \times (\mathbf{r'}-\mathbf{r})/\abs{\mathbf{r'}-\mathbf{r}}$ is the in-plane DMI with strength $D$ between nearest-neighbor lattice sites at positions $\mathbf{r}$ and $\mathbf{r'}$, and $B$ is the external magnetic field perpendicular to the plane. In addition, the first summation runs over all unique nearest-neighbor spin pairs. We fix $J=1$ so that $J$ sets the energy scale throughout the rest of this article.


\definecolor{greenB}{RGB}{64, 158, 0}
\definecolor{redA}{RGB}{222, 45, 38}
\usetikzlibrary {arrows.meta, bending} 
\newcommand*\rows{2}
\newcommand*\lattspacing{1.5}

\begin{figure}
    \centering
    

    \begin{minipage}[b][][s]{0.64\linewidth}
    \centering
    \begin{tikzpicture}
    \centering
    \foreach \row in {0, 1, ..., \rows} {
    \draw ($\row*(\lattspacing*0.5, {\lattspacing*0.5*sqrt(3)})$) -- ($(\rows*\lattspacing,0)+\row*(\lattspacing*0.5, {\lattspacing*0.5*sqrt(3)})$);
    \draw ($(\lattspacing*\row, 0)$) -- ($(\row*\lattspacing,0)+\rows*(0.5*\lattspacing, {0.5*\lattspacing*sqrt(3)})$);
    }
    \draw ($(0.5*\lattspacing, {0.5*\lattspacing*sqrt(3)})$) -- ($(\lattspacing, 0)$);
    \draw ($2*(0.5*\lattspacing, {0.5*\lattspacing*sqrt(3)})$) -- ($(2*\lattspacing, 0)$);
    \draw ($2*(0.5*\lattspacing, {0.5*\lattspacing*sqrt(3)}) + (\lattspacing,0)$) -- ($(2*\lattspacing,0)+ (0.5*\lattspacing,{0.5*\lattspacing*sqrt(3)})$);
    \filldraw[red] ({0.5*\lattspacing},{\lattspacing*0.5*sqrt(3)}) circle (2pt) node[anchor=east]{A};
    \filldraw[greenB] (0,0) circle (2pt) node[anchor=north]{B};
    \filldraw[blue] (\lattspacing,0) circle (2pt) node[anchor=north]{C};
    \filldraw[greenB] ({1.5*\lattspacing},{\lattspacing*0.5*sqrt(3)}) circle (2pt) node[anchor=north]{};
    \filldraw[blue] (\lattspacing,{\lattspacing*sqrt(3)}) circle (2pt) node[anchor=north]{};
    \filldraw[red] ({2*\lattspacing},0) circle (2pt) node[anchor=east]{};
    \filldraw[red] ({2*\lattspacing},{\lattspacing*sqrt(3)}) circle (2pt) node[anchor=east]{};
    \filldraw[blue] ({2.5*\lattspacing},{\lattspacing*0.5*sqrt(3)}) circle (2pt) node[anchor=east]{};
    \filldraw[greenB] ({3*\lattspacing},{\lattspacing*sqrt(3)}) circle (2pt) node[anchor=north]{};

    \draw [thick, arrows = {-Stealth[length=6pt]}] ({0.5*\lattspacing}, 0.15) -- ({0.5*\lattspacing}, {-0.25*\lattspacing});
    \draw [thick, arrows = {-Stealth[length=6pt]}] ({0.5*\lattspacing}, -0.15) -- ({0.5*\lattspacing}, {0.25*\lattspacing});
    \draw [thick, arrows = {-Stealth[length=6pt]}] ({1.5*\lattspacing}, 0.15) -- ({1.5*\lattspacing}, {-0.25*\lattspacing});
    \draw [thick, arrows = {-Stealth[length=6pt]}] ({1.5*\lattspacing}, -0.15) -- ({1.5*\lattspacing}, {0.25*\lattspacing});

    \draw[thick, arrows = {-Stealth[length=6pt]}] (\lattspacing, {-0.15*\lattspacing+0.5*sqrt(3)*\lattspacing}) -- (\lattspacing, {0.25*\lattspacing+0.5*sqrt(3)*\lattspacing});
    \draw[thick, arrows = {-Stealth[length=6pt]}] (\lattspacing, {0.15*\lattspacing+0.5*sqrt(3)*\lattspacing}) -- (\lattspacing, {-0.25*\lattspacing+0.5*sqrt(3)*\lattspacing});
    \draw[thick, arrows = {-Stealth[length=6pt]}] ({2*\lattspacing}, {-0.15*\lattspacing+0.5*sqrt(3)*\lattspacing}) -- ({2*\lattspacing}, {0.25*\lattspacing+0.5*sqrt(3)*\lattspacing});
    \draw[thick, arrows = {-Stealth[length=6pt]}] ({2*\lattspacing}, {0.15*\lattspacing+0.5*sqrt(3)*\lattspacing}) -- ({2*\lattspacing}, {-0.25*\lattspacing+0.5*sqrt(3)*\lattspacing});

    \draw[thick, arrows = {-Stealth[length=6pt]}] ({1.5*\lattspacing}, {sqrt(3)*\lattspacing}) -- ({1.5*\lattspacing}, {0.25*\lattspacing+sqrt(3)*\lattspacing});
    \draw[thick, arrows = {-Stealth[length=6pt]}] ({1.5*\lattspacing}, {sqrt(3)*\lattspacing}) -- ({1.5*\lattspacing}, {-0.25*\lattspacing+sqrt(3)*\lattspacing});
    \draw[thick, arrows = {-Stealth[length=6pt]}] ({2.5*\lattspacing}, {sqrt(3)*\lattspacing}) -- ({2.5*\lattspacing}, {0.25*\lattspacing+sqrt(3)*\lattspacing});
    \draw[thick, arrows = {-Stealth[length=6pt]}] ({2.5*\lattspacing}, {0.15*\lattspacing+sqrt(3)*\lattspacing}) -- ({2.5*\lattspacing}, {-0.25*\lattspacing+sqrt(3)*\lattspacing});

    \draw[thick, arrows = {-Stealth[length=6pt]}] ({0.25*\lattspacing}, {sqrt(3)/4*\lattspacing}) -- ($({0.25*\lattspacing}, {\lattspacing*sqrt(3)/4}) + 0.25*({0.5*sqrt(3)*\lattspacing}, {-0.5*\lattspacing})$);
    \draw[thick, arrows = {-Stealth[length=6pt]}] ({0.25*\lattspacing}, {sqrt(3)/4*\lattspacing}) -- ($({0.25*\lattspacing}, {\lattspacing*sqrt(3)/4}) - 0.25*({0.5*sqrt(3)*\lattspacing}, {-0.5*\lattspacing})$);
    \draw[thick, arrows = {-Stealth[length=6pt]}] ({0.75*\lattspacing}, {\lattspacing*3*sqrt(3)/4}) -- ($({0.75*\lattspacing}, {3*sqrt(3)/4*\lattspacing}) + 0.25*({0.5*sqrt(3)*\lattspacing}, {-0.5*\lattspacing})$);
    \draw[thick, arrows = {-Stealth[length=6pt]}] ({0.75*\lattspacing}, {3*sqrt(3)/4*\lattspacing}) -- ($({0.75*\lattspacing}, {\lattspacing*3*sqrt(3)/4}) - 0.25*({0.5*sqrt(3)*\lattspacing}, {-0.5*\lattspacing})$);

    \draw[thick, arrows = {-Stealth[length=6pt]}] ({1.25*\lattspacing}, {\lattspacing*sqrt(3)/4}) -- ($({1.25*\lattspacing}, {\lattspacing*sqrt(3)/4}) + 0.25*({0.5*sqrt(3)*\lattspacing}, {-0.5*\lattspacing})$);
    \draw[thick, arrows = {-Stealth[length=6pt]}] ({1.25*\lattspacing}, {\lattspacing*sqrt(3)/4}) -- ($({1.25*\lattspacing}, {\lattspacing*sqrt(3)/4}) - 0.25*({0.5*sqrt(3)*\lattspacing}, {-0.5*\lattspacing})$);
    \draw[thick, arrows = {-Stealth[length=6pt]}] ({1.75*\lattspacing}, {3*sqrt(3)/4*\lattspacing}) -- ($({1.75*\lattspacing}, {\lattspacing*3*sqrt(3)/4}) + 0.25*({0.5*sqrt(3)*\lattspacing}, {-0.5*\lattspacing})$);
    \draw[thick, arrows = {-Stealth[length=6pt]}] ({1.75*\lattspacing}, {\lattspacing*3*sqrt(3)/4}) -- ($({1.75*\lattspacing}, {\lattspacing*3*sqrt(3)/4}) - 0.25*({0.5*sqrt(3)*\lattspacing}, {-0.5*\lattspacing})$);

    \draw[thick, arrows = {-Stealth[length=6pt]}] ({2.25*\lattspacing}, {\lattspacing*sqrt(3)/4}) -- ($({2.25*\lattspacing}, {\lattspacing*sqrt(3)/4}) + 0.25*({0.5*sqrt(3)*\lattspacing}, {-0.5*\lattspacing})$);
    \draw[thick, arrows = {-Stealth[length=6pt]}] ({2.25*\lattspacing}, {\lattspacing*sqrt(3)/4}) -- ($({2.25*\lattspacing}, {\lattspacing*sqrt(3)/4}) - 0.25*({0.5*sqrt(3)*\lattspacing}, {-0.5*\lattspacing})$);
    \draw[thick, arrows = {-Stealth[length=6pt]}] ({2.75*\lattspacing}, {\lattspacing*3*sqrt(3)/4}) -- ($({2.75*\lattspacing}, {\lattspacing*3*sqrt(3)/4}) + 0.25*({0.5*sqrt(3)*\lattspacing}, {-0.5*\lattspacing})$);
    \draw[thick, arrows = {-Stealth[length=6pt]}] ({2.75*\lattspacing}, {\lattspacing*3*sqrt(3)/4}) -- ($({2.75*\lattspacing}, {\lattspacing*3*sqrt(3)/4}) - 0.25*({0.5*sqrt(3)*\lattspacing}, {-0.5*\lattspacing})$);

    \draw[thick, arrows = {-Stealth[length=6pt]}] 
    ({0.75*\lattspacing}, {\lattspacing*sqrt(3)/4}) -- ($({0.75*\lattspacing}, {\lattspacing*sqrt(3)/4}) + 0.25*({0.5*sqrt(3)*\lattspacing}, {0.5*\lattspacing})$);
    \draw[thick, arrows = {-Stealth[length=6pt]}] ({0.75*\lattspacing}, {\lattspacing*sqrt(3)/4}) -- ($({0.75*\lattspacing}, {\lattspacing*sqrt(3)/4}) - 0.25*({0.5*sqrt(3)*\lattspacing}, {0.5*\lattspacing})$);
    \draw[thick, arrows = {-Stealth[length=6pt]}] 
    ({1.25*\lattspacing}, {\lattspacing*3*sqrt(3)/4}) -- ($({1.25*\lattspacing}, {\lattspacing*3*sqrt(3)/4}) + 0.25*({0.5*sqrt(3)*\lattspacing}, {0.5*\lattspacing})$);
    \draw[thick, arrows = {-Stealth[length=6pt]}] ({1.25*\lattspacing}, {3*sqrt(3)/4*\lattspacing}) -- ($({1.25*\lattspacing}, {\lattspacing*3*sqrt(3)/4}) - 0.25*({0.5*sqrt(3)*\lattspacing}, {0.5*\lattspacing})$);

    \draw[thick, arrows = {-Stealth[length=6pt]}] 
    ({1.75*\lattspacing}, {\lattspacing*sqrt(3)/4}) -- ($({1.75*\lattspacing}, {\lattspacing*sqrt(3)/4}) + 0.25*({0.5*sqrt(3)*\lattspacing}, {0.5*\lattspacing})$);
    \draw[thick, arrows = {-Stealth[length=6pt]}] ({1.75*\lattspacing}, {\lattspacing*sqrt(3)/4}) -- ($({1.75*\lattspacing}, {\lattspacing*sqrt(3)/4}) - 0.25*({0.5*sqrt(3)*\lattspacing}, {0.5*\lattspacing})$);
    \draw[thick, arrows = {-Stealth[length=6pt]}] 
    ({2.25*\lattspacing}, {\lattspacing*3*sqrt(3)/4}) -- ($({2.25*\lattspacing}, {\lattspacing*3*sqrt(3)/4}) + 0.25*({0.5*sqrt(3)*\lattspacing}, {0.5*\lattspacing})$);
    \draw[thick, arrows = {-Stealth[length=6pt]}] ({2.25*\lattspacing}, {\lattspacing*3*sqrt(3)/4}) -- ($({2.25*\lattspacing}, {3*sqrt(3)/4*\lattspacing}) - 0.25*({0.5*sqrt(3)*\lattspacing}, {0.5*\lattspacing})$);
    
    \end{tikzpicture}
    \end{minipage}
    \hfill
    \begin{minipage}[b][][s]{0.34\linewidth}
    \centering
    \begin{tikzpicture}
    \centering
    \draw ($(1, 0)$) -- ($(2, 0)$);
    \draw ($(1, 0)$) -- ($(1,0)+\rows*(0.5, {0.5*sqrt(3)})$);
    \draw ($(0.5, {0.5*sqrt(3)})$) -- ($(2.5, {0.5*sqrt(3)})$);
    \draw ($2*(0.5, {0.5*sqrt(3)})$) -- ($(1,0)+2*(0.5, {0.5*sqrt(3)})$);
    \draw ($(0.5, {0.5*sqrt(3)})$) -- ($(1, 0)$);
    \draw ($2*(0.5, {0.5*sqrt(3)})$) -- ($(2, 0)$);
    \draw ($2*(0.5, {0.5*sqrt(3)}) + (1,0)$) -- ($(2,0)+ (0.5,{0.5*sqrt(3)})$);
    \draw (0.5,{0.5*sqrt(3)}) -- (1,{sqrt(3)});
    \draw (2,0) -- (2.5,{0.5*sqrt(3)});
    \filldraw[red] (0.5,{0.5*sqrt(3)}) circle (2pt) node[anchor=east]{{\color{black}$j$}};
    \filldraw[blue] (1,0) circle (2pt) node[anchor=north]{{\color{black}$k$}};
    \filldraw[greenB] (1.5,{0.5*sqrt(3)}) circle (2pt) node[anchor=south]{{\color{black}$i$}};
    \filldraw[blue] (1,{sqrt(3)}) circle (2pt) node[anchor=north]{};
    \filldraw[red] (2,0) circle (2pt) node[anchor=east]{};
    \filldraw[red] (2,{sqrt(3)}) circle (2pt) node[anchor=south]{{\color{black}$k$}};
    \filldraw[blue] (2.5,{0.5*sqrt(3)}) circle (2pt) node[anchor=west]{{\color{black}$j$}};
    \draw[thick, -{[flex']>}] (1.19, {1/sqrt(3)}) arc (0:300:0.19cm); 
    \draw[thick, -{[flex']>}] (2.19, {2/sqrt(3)}) arc (0:300:0.19cm);
    
    \end{tikzpicture}
    \end{minipage}
    
    \caption{Left: Sites that belong to each of the three sublattices of the triangular lattice, where ``A'', ``B'', ``C'' label spin sites in each of the sublattices. Sites having the same color form part of the same sublattice. Arrows indicate the directions of the DMI vectors $\mathbf{D}_{\mathbf{r'},\mathbf{r}}$.  Right: Two representative sets of sites $(i,j,k)$ used to compute the chirality in Eq.\eqref{eq:quir}.  
    We label the sites of the vertices of each triangular plaquette as $i$, $j$ and $k$ following a counterclockwise order. Note that each site belongs to a different sublattice. }
    \label{fig:sitios_quir}
\end{figure}
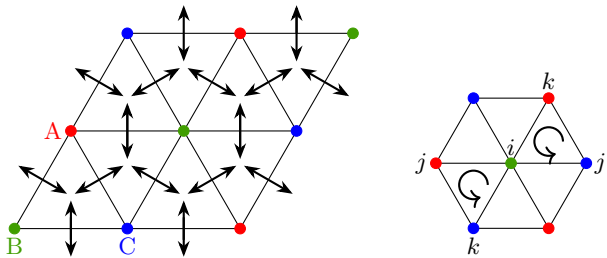

The classical counterpart of the Hamiltonian from Eq.(\ref{eq:H}) exhibits an antiferromagnetic skyrmion lattice phase, stabilized in a broad range of temperatures and intermediate magnetic fields \cite{diegoAFskys,Zukovic1}. This phase is formed by three interpenetrated ferromagnetic skyrmion lattices, one in each sublattice of the triangular lattice (sublattices defined in Fig.~\ref{fig:sitios_quir}). This type of texture has been shown to be experimentally relevant to MnSc$_2$S$_4$ \cite{Gao2020,Rosales2022} and similar phases are known to arise in other frustrated models \cite{mohylna2022spontaneous2,Zukovic2,mohylna2025}. At lower fields, interpenetrated helices form an antiferromagnetic helices pattern, which leads to the antiferromagnetic skyrmion lattice as the magnetic field is increased. For larger magnetic fields, intermediate skyrmion-like textures arise before reaching saturation. 

In the present work, we take $S=1/2$ and resort to DMRG to study signatures of quantum antiferromagnetic skyrmions. We used the DMRG implementation from the ITensorMPS library, which is based on the Julia version of ITensor \cite{itensorpaper}, to perform our calculations. See the Appendix \ref{ap:convergence} for information on convergence and technical details. 

To analyze the ground states of $\hat{H}$ from Eq.\eqref{eq:H}, we worked with lattices of {$N=L\times L$ sites with $L=12-20$}, taking open boundary conditions (OBC) and fixing $D/J=0.8$. We focus especially at intermediate magnetic fields, where skyrmion-hosting phases may arise. We find evidence of three-sublattice skyrmion textures for a wide range of magnetic fields and irrespective of how the boundaries of the lattice are chosen, as we will show in the next section. We also present magnetization profiles and compute the chirality, which is {the calculation of the topological charge $Q$ in a discrete lattice}. In classical systems, a non-zero chirality is associated with skyrmions. This has also been the case for quantum ferromagnetic skyrmions hosted in spin-1/2 systems \cite{sky_ED}. In addition, we inspect the quantum entanglement of these antiferromagnetic textures, which is a strictly non-classical phenomenon. We discuss our findings in the next section.

\section{Results}
\label{sec:results}

As stated before, the main aim of this work is to identify the different emerging phases for the Hamiltonian in Eq.(\ref{eq:H}) for $S=1/2$ spins. For that purpose, we computed the magnetization and scalar chirality of the system and found three main phases: antiferromagnetic helical phases at lower fields, an intermediate antiferromagnetic skyrmion phase and finally, a fully polarized phase.  As in the classical model, there is also an intermediate chiral phase as skyrmions become polarized and before the system goes into the ferromagnetic phase. We also explore the emergent skyrmion phase in reciprocal space calculating the structure factor. Furthermore, despite the expected boundary effects due to the system size and OBC, we show that the antiferromagnetic skyrmion texture is found for different shapes of the lattice boundaries and the scale of the skyrmions in the bulk changes with the DMI strength. Then, we explore the quantum nature of the resulting ground states studying their  entanglement entropy and concurrence.

\begin{figure}[h!]
    \includegraphics[width=\linewidth]{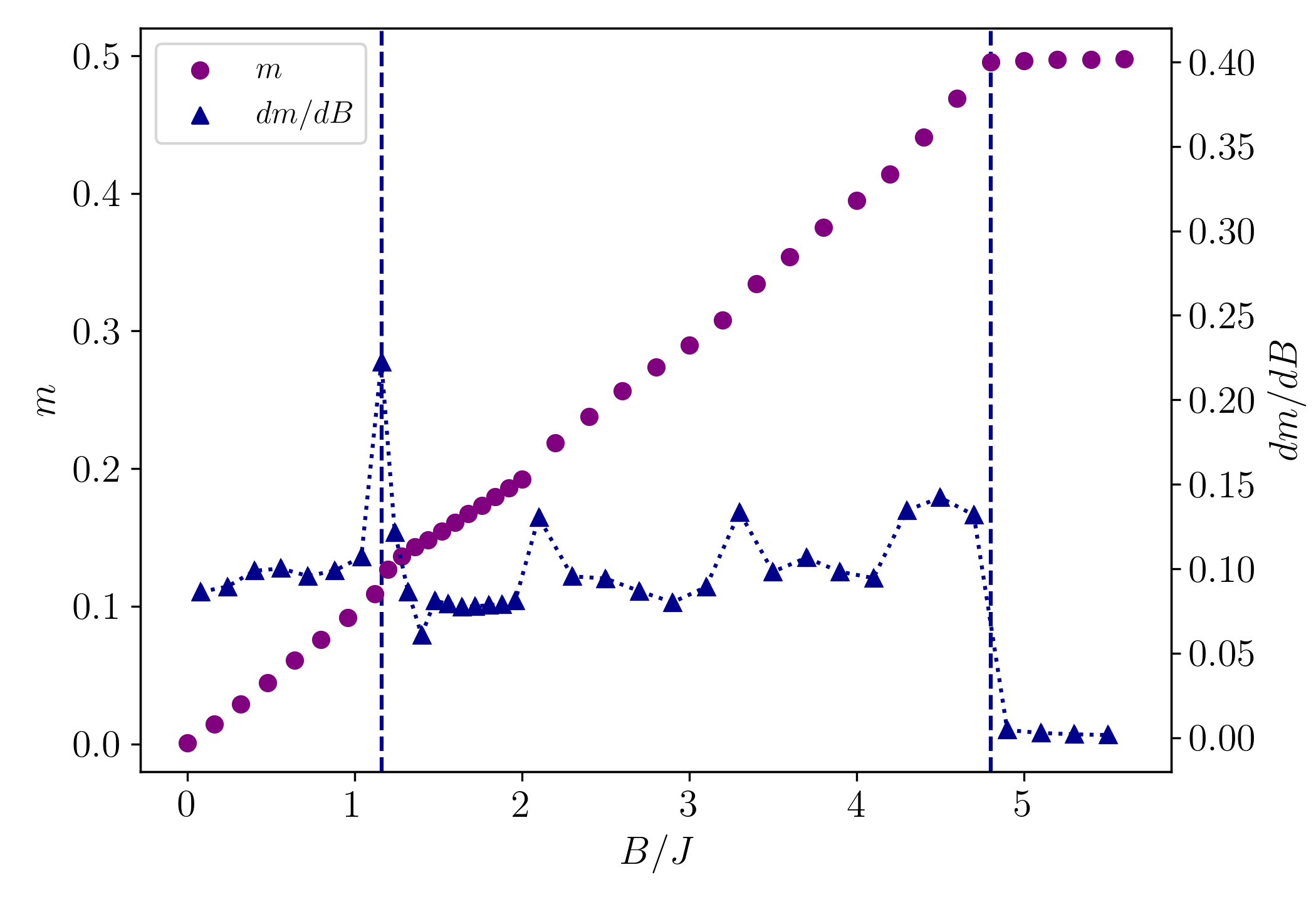}
    \caption{Average magnetization per site $m$ in the magnetic field direction (purple circles) and its derivative, the susceptibility (dark blue triangles), as a function of magnetic field $B/J$. As expected, $m$ approaches the value $1/2$ for high $B/J$. The susceptibility is discontinuous at $B/J \simeq 1.16$ and $B/J \simeq 4.8$ (dashed dark blue vertical lines). 
    }
    \label{magn}
\end{figure}

\begin{figure*}[ht!]
\includegraphics[width=\textwidth]{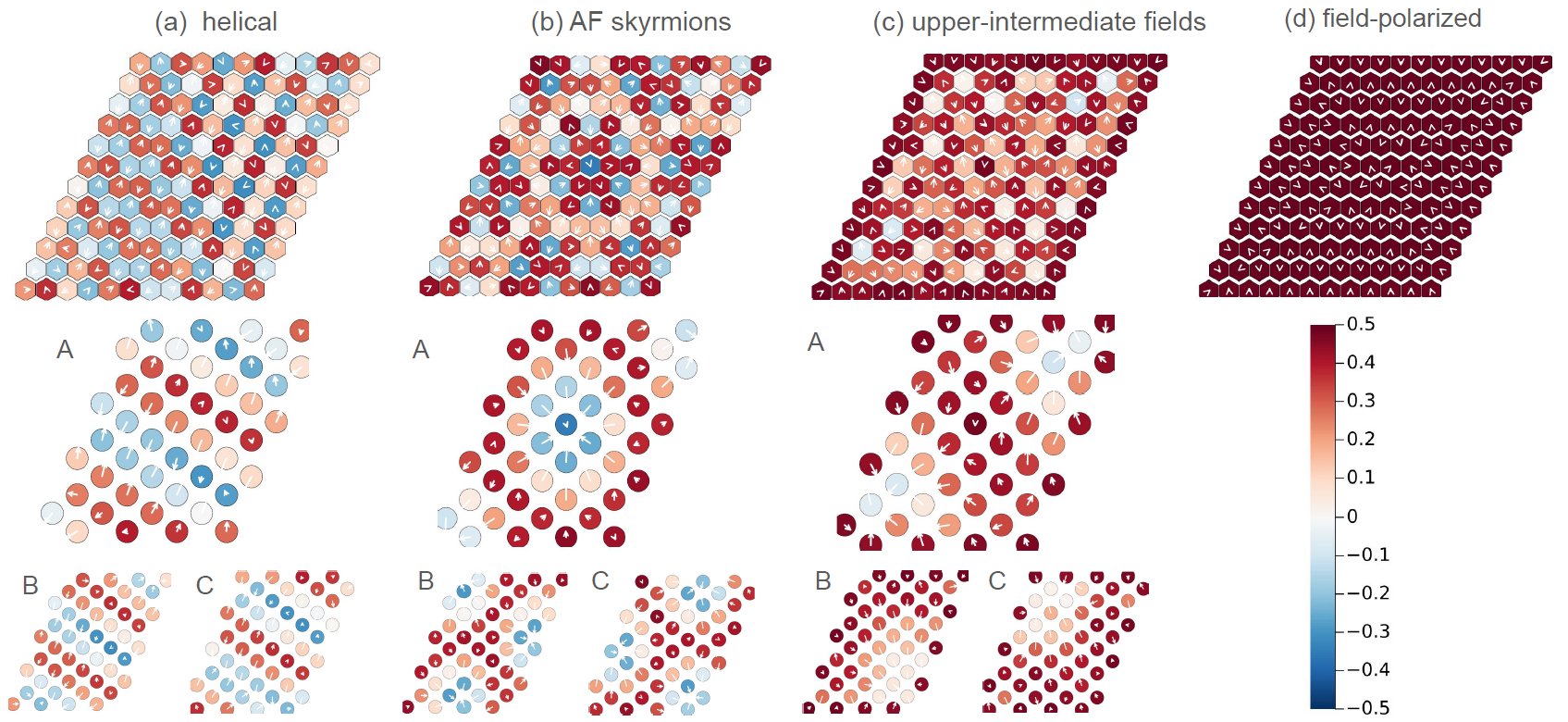}
    \caption{Representative magnetization profiles for the obtained ground states: (a) AF helical phase  at $B/J=0.8$, (b) AF skyrmions at $B/J=1.76$, (c) higher-field AF skyrmion-like phase at $B/J=3.4$, and (d) field-polarized at $B/J=5.4$. Local magnetization of the entire lattice for each phase is presented on the first row. For the first three cases, we also display the profiles for each triangular sublattice (A, B and C, as in Fig.~\ref{fig:sitios_quir}). The colorbar indicates the $\langle \hat{S}^z\rangle$ scale.}
   
    \label{fig:Mz_variosB}
\end{figure*}

\begin{figure}[h!]
\includegraphics[width=0.8\linewidth]{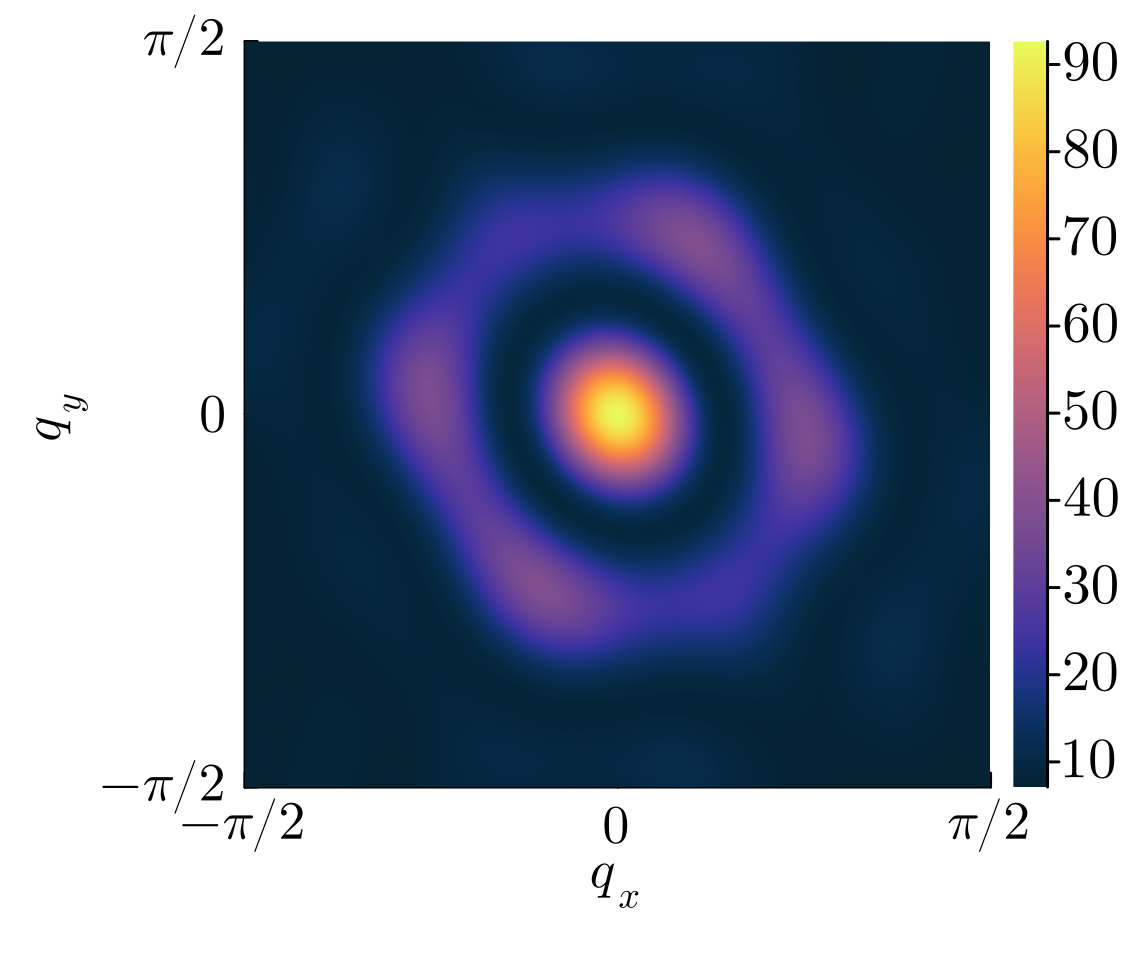}
    \centering
    \caption{Structure factor $\mathcal{S}_{zz}$ (in arbitrary units) of sublattice A from Fig.~\ref{fig:Mz_variosB}(b). It resembles $\mathcal{S}_{zz}$ of a lattice hosting a single quantum F skyrmion from Ref. \cite{Qsky}.}
\label{fig:Sq}
\end{figure}

\subsection{Magnetization}
\label{sec:magn}
We begin our study by computing expectation values of local spin operators. Magnetization per site $m=\left( \sum_{\mathbf{r}}\langle \hat{S}^z_{\mathbf{r}}\rangle \right)/N$ for each value of $B/J$ is presented in Fig.~\ref{magn}. From now on, angle brackets $\langle \rangle$ denote the expectation value of the corresponding operator (in this case, of $\hat{S}^z$ at site $\mathbf{r}$). As we can see, it reaches its maximum value ($m=1/2$) for high magnetic fields. This means that the system is field-polarized for $B/J \gtrsim 4.8$.

At plain sight, $m$ seems to be continuous for the full range of $B/J$. To search for signatures of possible different phases, we compute its derivative (i.e. the magnetic susceptibility $\frac{dm}{dB}$). Susceptibility shows a peak at $B/J \simeq 1.16$ and a jump at $B/J \simeq 4.8$. This indicates that $m$ has a single jump discontinuity around $B/J \simeq 1.16$ unlike the ferromagnetic system, where clear jumps in $m$ delimit the skyrmion region \cite{Qsky}. These changes in the behavior of $\frac{dm}{dB}$ with $B$ are signatures of the stabilization of different phases, as we confirm with other numerical results in the next subsections.

As we mentioned above, there seem to be three main regions demarked by discontinuities in the susceptibility. We turn to the local magnetization of the ground states to check if this is the case. We show the real-space magnetization profiles of the system, i.e. $\langle \hat{\mathbf{S}}_i \rangle \equiv (\langle \hat{S}^x_i\rangle,\langle \hat{S}_i ^y\rangle,\langle \hat{S}^z_i\rangle)$ for each spin $i$, at different values of $B/J$ in Fig.~\ref{fig:Mz_variosB}. The first and second components of $\langle \hat{\mathbf{S}}_i \rangle$ are represented by arrows, and colors denote spin expectation values $\langle \hat{S}^z_i\rangle$.

The low field phase is depicted in panel (a) of Fig.~\ref{fig:Mz_variosB}: an antiferromagnetic helical phase formed by three interpenetrated helices, one in each triangular sublattice (A, B, C). The antiferromagnetic skyrmion phase formed by three interpenetrated ferromagnetic skyrmion textures is shown in panel (b), corresponding to the intermediate field region.  We notice however that the magnetization patterns found at higher field values of the intermediate region differ markedly from those of panel (b): panel (c) shows these patterns, which are non-trivial skyrmion-like textures {that do not have fully polarized cores} similar to those found in the classical case \cite{diegoAFskys}. We will further discuss this in the next subsection, where we analyze the scalar chirality. The third region, where $m\simeq1/2$ corresponds to the field-polarized phase, and is illustrated in panel (d).

\begin{figure}[h!]
    \includegraphics[width=\linewidth]{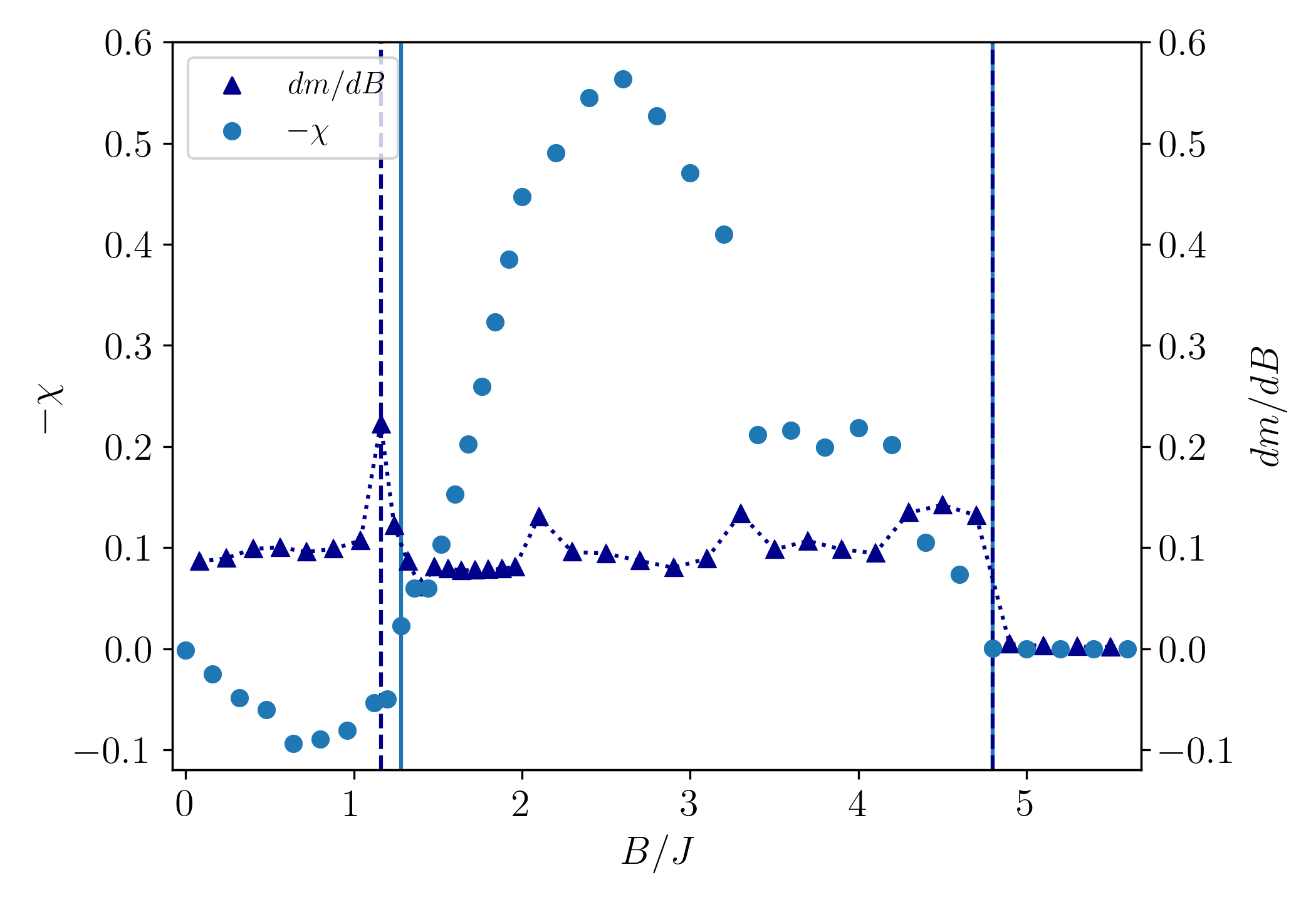}
    \centering
    \caption{Chirality {$\chi$} (blue circles) and magnetic susceptibility (dark blue triangles) as a function of $B/J$. We multiply {$\chi$} by $-1$ so that {$\chi$} and the susceptibility take mostly positive values, which allows for an easier comparison of changes in their behavior. Solid blue lines indicate $B/J$ values where {$\chi$} changes sign {at lower fields and goes to zero at higher fields, }and dashed dark blue lines mark discontinuities in the susceptibility. Note that at higher fields both lines show up at $B/J\simeq4.8$. } 
    
    \label{fig:Q_chi}
\end{figure}

\begin{figure}[h!]
    \includegraphics[width=\linewidth]{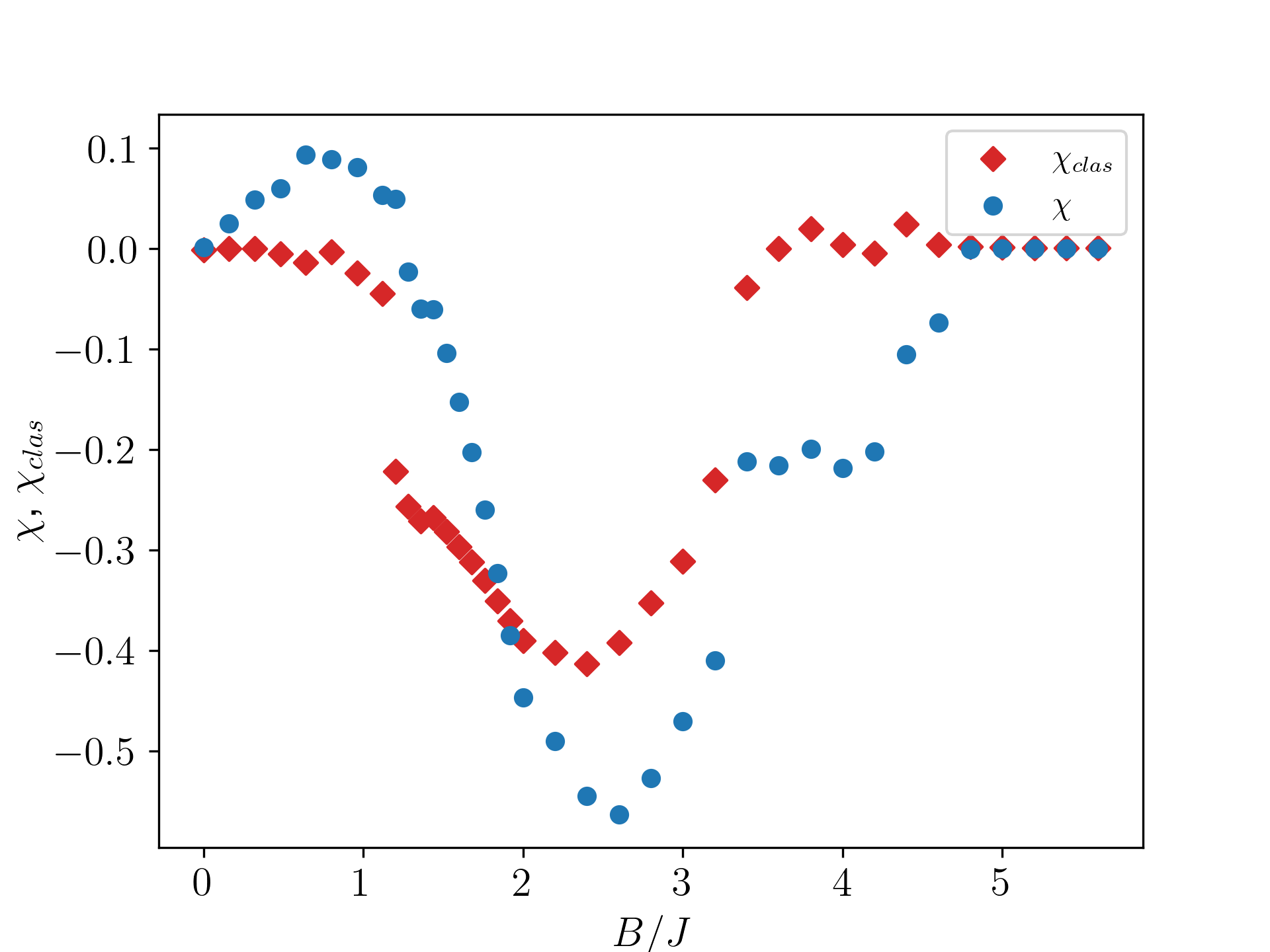}
    \centering
    \caption{Classical chirality {$\chi_{clas}$} (red diamonds) and quantum chirality {$\chi$} (blue circles) as a function of $B/J$.} 
    
    \label{fig:Qclas}
\end{figure}
\begin{figure*}
    \centering
    \includegraphics[width=0.95\textwidth]{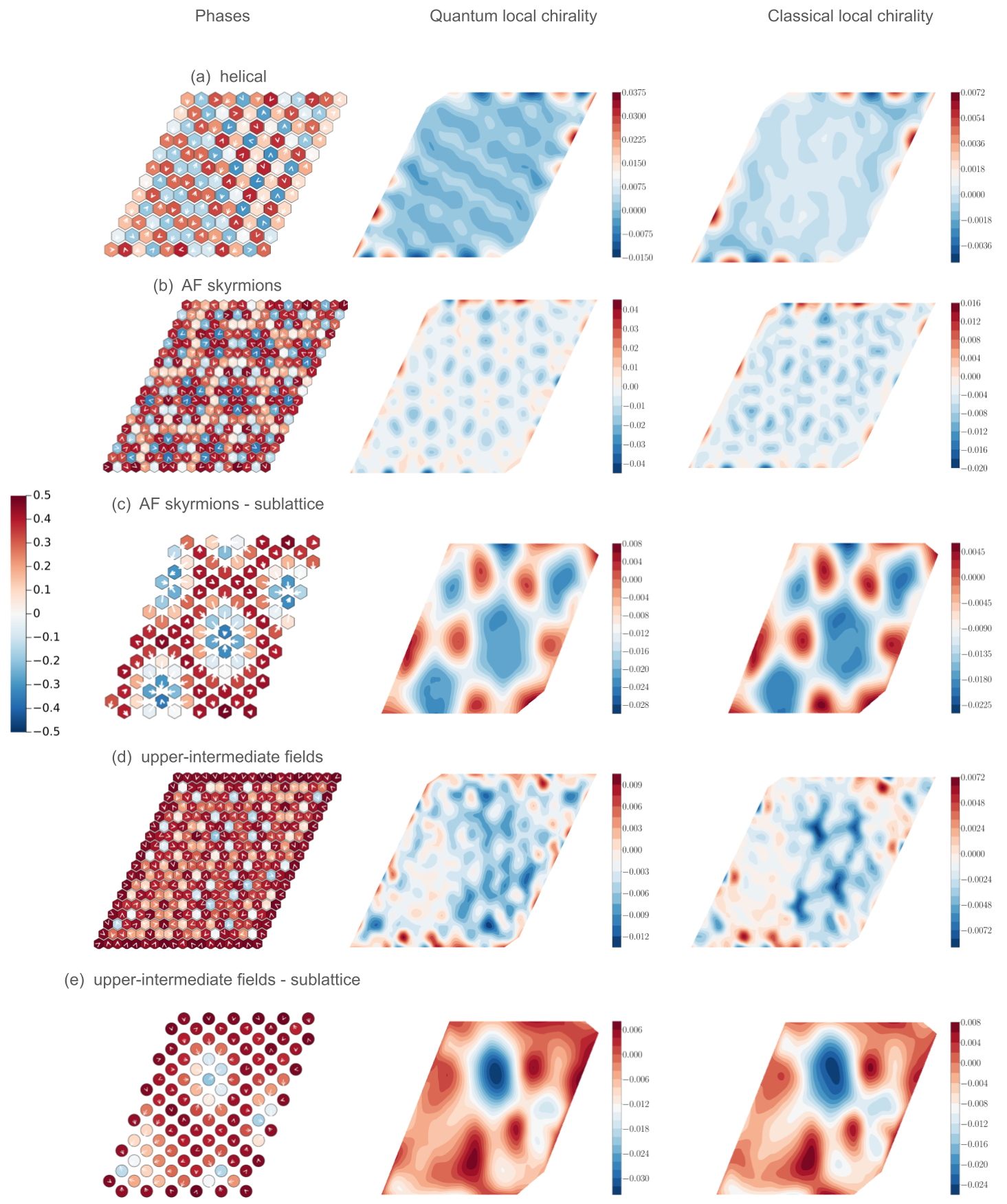}
    \caption{{Quantum (middle column) and classical (right column) local chirality maps for different phases: helical ( (a), $L=12$), antiferromagnetic skyrmions ( (b) and (c), $L=18$), higher-field skyrmion-like phase ( (d) and (e), $L=18$). Corresponding magnetization profiles and their common color bar for $\langle S^z\rangle$ are presented in the left column. Notice that each local chirality map has its own colorbar.} 
    }
    \label{fig:chilocal}
\end{figure*}

\subsection{{Structure factor}}
\label{sec:Sq}

 So far we have {seen} that non-trivial magnetic spin textures emerge as ground states of the simple antiferromagnetic model introduced in Eq.\eqref{eq:H} at intermediate magnetic fields.  Based on their magnetization patterns, these textures {arising at intermediate magnetic fields }can be regarded to be antiferromagnetic skyrmion textures. The magnetization pattern of sublattice A from Fig.~\ref{fig:Mz_variosB}(b) resembles that of a quantum ferromagnetic skyrmion. We then resort to the spin-spin correlation function to check if sublattice spins are also correlated with each other in the same way as spins from quantum ferromagnetic skyrmions. We compute the structure factor
$$\mathcal{S}_{\alpha \beta}(\mathbf{q})=\sum_{\mathbf{r},\mathbf{r}'} e^{i\mathbf{q}(\mathbf{r} - \mathbf{r}')} \langle\hat{S}_{\mathbf{r}}^{\alpha}\hat{S}_{\mathbf{r}'}^{\beta}\rangle,$$
which is the Fourier transform of the spin-spin correlation function between spins at positions $\mathbf{r}$ and $\mathbf{r'}$, with $\alpha, \beta = x,y,z$. We are interested in the sublattice structure factor, so the sum in $\mathcal{S}_{\alpha \beta}(\mathbf{q})$ runs over sites of the same sublattice.

Fig.~\ref{fig:Sq} shows $\mathcal{S}_{zz}$ of  sublattice A  of the ground state for $B/J=1.76$. We notice that the $\mathcal{S}_{zz}$ of the sublattice resembles the $\mathcal{S}_{zz}$ of a system hosting  a single quantum ferromagnetic skyrmion  \cite{Qsky}. There is a Bragg peak at $\mathbf{q}=0$ due to the non-zero magnetization in the $\hat{z}$ direction. This agrees with the three sublattice pattern found in the local magnetization (Fig.~\ref{fig:Mz_variosB}(b)) where, for this system size, a single ferromagnetic skyrmion pattern is realized in one of the sublattices.

\subsection{{Scalar chirality}}
\label{sec:chirality}

Scalar chirality is a commonly-used quantity that corresponds to {the calculation of the topological charge $Q$} for a discrete lattice. {Given that usually the radius of a skyrmion is of a few lattice spacings, effectively discretizing the calculation, the resulting total chirality does not exactly yield the total number of skyrmions. However, it strongly }signals the emergence of magnetic skyrmions{. It is the usual order parameter associated to non-trivial topological phases calculated} in classical spin systems \cite{Okubo2012}. On a triangular lattice, it is defined as
\begin{equation}
   {\chi_{latt}}=\frac{1}{8\pi}  \sum_{(i,j,k)\:\in\:\Delta\:/\:\nabla} \mathbf{m}_i \cdot (\mathbf{m}_j \times \mathbf{m}_k),
\label{eq:Qclas}    
\end{equation}
where $\mathbf{m}_i$, $\mathbf{m}_j$ and $\mathbf{m}_k$ are classical magnetic moments of unit length located at sites $i$, $j$ and $k$ respectively. The sum runs over sites $i$, $j$ and $k$ that form each non-overlapping triangular plaquette. Examples of spins $i$, $j$ and $k$ on an upward triangular plaquette ($\Delta$) and on a downward one ($\nabla$) of the lattice are shown in Fig.~\ref{fig:sitios_quir}. 

\subsubsection{{Total Scalar Chirality}}

In what follows, we will apply a quantized version of the scalar chirality \cite{sky_ED}, {calculating the expectation value of the triple product of the spin operators at three sites forming a triangular plaquette,} given by

\begin{equation}
     {\chi}=\frac{1}{\pi}  \sum_{(i,j,k)\:\in\:\Delta\:/\:\nabla} \langle\mathbf{\hat{S}}_i \cdot (\mathbf{\hat{S}}_j \times \mathbf{\hat{S}}_k)\rangle,
    \label{eq:quir}    
\end{equation}
\noindent where the sum once again runs over sites $i$, $j$ and $k$ that form an elementary triangular plaquette (see Fig.~\ref{fig:sitios_quir}). We will refer to the quantum scalar chirality {$\chi$} as quantum chirality or chirality interchangeably going forward. 

Chirality as a function of magnetic field $B/J$ is shown in Fig.~\ref{fig:Q_chi}. At first sight, we see that the chirality {$\chi$} defines three main regions. At lower fields, {$\chi$} is comparatively small and positive up to $B/J\simeq 1.2$. Then there is a range of magnetic field values in which {$\chi$} is negative, and its absolute value {$|\chi|$ gets} visibly higher than for lower fields, reaching its maximum value at $B/J \simeq 2.6$. Finally, {$|\chi|$} decreases smoothly until it reaches 0 at $ B/J \simeq 4.8$.

As was done for the magnetization $m$, we compare the magnetic susceptibility with the chirality {$\chi$} in Fig.~\ref{fig:Q_chi}. As mentioned in the previous subsection, the magnetic susceptibility has a peak at $B/J \simeq 1.16$ and a jump at $B/J \simeq 4.8$. {For the chirality curves, we see that the first peak of the susceptibility at $B/J \simeq 1.16$ closely matches the value of the external field where $\chi$ changes sign and $|\chi|$ increases more sharply. The second jump in the susceptibility at higher fields agrees with the value of $B/J$ where $\chi$ drops to zero. Thus, we may establish three ranges of magnetic fields where different phases are expected.} At low magnetic fields ($B/J \lesssim 1.2$), a perfect helical phase would have {$\chi=0$}, and this is actually the case for $B/J=0$. However, {$|\chi|$} starts to increase with $B/J$. {This behavior is similar to that found in exact diagonalization studies of chirality for $S=1/2$ spins in a skyrmion} ferromagnetic model due to quantum fluctuations {\cite{sky_ED}}. Most importantly, {$\chi$ is non-zero in the central region ($1.2 \lesssim B/J \lesssim 4.8$), where the local magnetization patterns in Fig.~\ref{fig:Mz_variosB} indicate skyrmion and skyrmion-like textures are found}. {A detailed look into the behavior of the $\chi$ curve reveals that there are two subregions in this range of $B/J$}: one where {$|\chi|$} is significantly higher ($1.2 \lesssim B/J \lesssim 3.2$), corresponding to the antiferromagnetic skyrmion phase (Fig.~\ref{fig:Mz_variosB}(b)), and a higher-field sector ($3.2 \lesssim B/J \lesssim 4.8$) where {$|\chi|$} drops, signalling the intermediate skyrmion-like phases illustrated in Fig.~\ref{fig:Mz_variosB}(c). {In these phases, as the system is magnetized with increasing magnetic field, the skyrmion cores are no longer fully polarized, and thus $|\chi|$ is smaller}. Finally, as expected, {$\chi=0$} for the field-polarized phase. Therefore, as in classical systems, and even in this antiferromagnetic case, the {chirality defined as the expectation value of the triple product of three spins may be used} as an order parameter to indicate {possible regions with} non-trivial {topological} textures. {For classical spins in a simple ferromagnetic model, abrupt jumps in $\chi_{clas}$ are to be expected at very low temperatures. However, thermal fluctuations induce small distortions and intermediate phases with non-zero chirality, which render a softer and more continuous scalar chirality curve at higher temperatures \cite{Ezawa2011}. In this quantum case, correlations and possible boundary effects give a not so sharp $\chi$ curve \cite{sky_ED}.}

We stated above that a {negative value of $\chi$} at lower magnetic fields can be a good indicator of the rise of skyrmions even if {$\chi\approx0$}. As a complement to this study, we compute {what we define as the classical scalar chirality} {$\chi_{clas}$,} replacing each $\mathbf{m}$ in Eq.\eqref{eq:Qclas} with $\langle 2\hat{\mathbf{S}} \rangle$, as was done in previous works {for $S=1/2$ ferromagnetic skyrmion models} \cite{PhysRevB.108.094410, PhysRevResearch.4.023111}. {This version of the scalar chirality, where the triple product is calculated with the expectation values of the spin operators, has fewer correlation effects than the previously-defined quantum scalar chirality $\chi$.} Classical chirality {$\chi_{clas}$} vs. $B/J$ is presented in Fig.~\ref{fig:Qclas} along with {$\chi$}. It can be seen that {$\chi_{clas}$} is non-vanishing for $1.2 \lesssim B/J \lesssim 3.2$ (region where ground states display skyrmion patterns in their local magnetization) and  negligible elsewhere. {In fact, $\chi_{clas}$ shows two clear jumps delimiting the skyrmion region, in contrast to the more continuous $\chi$ curve dominated by quantum effects. $\chi_{clas}$ is close to zero in the helical region and in the skyrmion-like subphase, where in the latter case the spins are more polarized with the magnetic field. }Thus, {$\chi_{clas}$ and $\chi$} can be used to detect {these topological textures for $S=1/2$ spins, whether ferromagnetic, as shown previously in Refs. \cite{PhysRevB.108.094410, PhysRevResearch.4.023111}, or antiferromagnetic, as shown here.} However, contrary to what happens in the ferromagnetic case \cite{SkyF_SW}, quantum fluctuations suppress the lower-field chiral region with respect to the classical chirality diagram for this Hamiltonian from Ref.~\cite{spinwaves}. This is consistent with previous spin wave studies for these systems \cite{spinwaves, tesinaTome}.

{One open question is whether the value of $\chi$ is related to the stability of AF skyrmions, as was observed in the ferromagnetic case \cite{PhysRevResearch.4.023111}. Similarly, here we also find skyrmion textures with $\chi \approx 0$, which, upon comparison with $\chi_{clas}$, seems to be an effect of quantum fluctuations. This question will be addressed in future work.} 

\subsubsection{ {Local Scalar Chirality}}
 
 {The comparison between $\chi$ and $\chi_{clas}$ supports our claim about the main phases of the proposed model. To provide further validation, we will explore the local chirality maps for both $\chi$ and $\chi_{clas}$. The local chirality distribution in the lattice helps to determine the nature of the emergent texture. For example, it is particularly useful in meron-antimeron phases  \cite{Lin2015,mohylna2025,Wang2021}, where the meron-antimeron structures have opposite topological charge (and thus, opposite scalar chirality) that cancels out when considering the total lattice. For a system of classical spins, the local chirality at site $i$ is defined as $\displaystyle \chi_{latt}(i)=\frac{1}{8\pi} \sum_{\Delta,\nabla /\: i\in\{\Delta,\nabla\}}\mathbf{m}_i \cdot (\mathbf{m}_j \times \mathbf{m}_k)$, taking the triple product over the neighboring spins. Here we work as we did with $\chi$ and $\chi_{clas}$, replacing $\mathbf{m} \xrightarrow[]{} 2\hat{\mathbf{S}}$ in $\chi_{latt}(i)$ and calculating the expectation value of the triple product in one case (what we will from now on refer to as $\chi(i)$), and replacing $\mathbf{m} \xrightarrow[]{}\langle 2\hat{\mathbf{S}}\rangle$ in the other ($\chi_{clas}(i)$).
 
 Representative magnetization profiles and their corresponding quantum and classical chirality  maps are presented in Fig.~\ref{fig:chilocal}. Inspection of  these patterns reinforces our findings for each region:}

{
 \begin{itemize}
 \item{\textit{Helical phase:} In the chirality maps from Fig.~\ref{fig:chilocal}(a), we can see that the  positive total chirality seen at low fields in Fig.~\ref{fig:Q_chi} is mainly an effect of open boundary conditions: positive chiral regions are found in the corners of the lattice, whereas at the center the chirality values are close to zero, as expected in a helical phase in a classical system.}

 \item{\textit{Skyrmion phase:} The local chirality map for the complete lattice and for one sublattice in the skyrmion region ($B=1.76$) are shown in panels Fig.~\ref{fig:chilocal}(b) and (c), respectively. Both $\chi(i)$ and $\chi_{clas}(i)$ maps of the entire lattice show periodically distributed regions with negative values. The local sublattice chirality confirms the presence of triangular skyrmion sublattices: the negative chirality matches the skyrmion core, and is surrounded by regions with $\chi\approx0$, corresponding to the ferromagnetic background. Positive chirality regions are located at the borders and arise due to the open boundary effects.}

 \item{\textit{Skyrmion-like subphase:} Observation of the local chirality parameters for the higher field region (Fig.~\ref{fig:chilocal}(d) for $B=3.4$) shows that the chirality values have lower absolute values in this phase with not-fully-polarized skyrmion cores, compared to the skyrmion  phase. Moreover, the distribution of $\chi(i)$ and $\chi_{clas}(i)$ does not seem as even as in the skyrmion phase. In  Fig.~\ref{fig:chilocal}(e) we present these results for one sublattice. In this case, the skyrmion-like structures on the sublattice are more separated since the surroundings are more polarized. Even after considering a system size larger than that from Fig.~\ref{fig:Mz_variosB} ($L=18$ vs. $L=12$),  only one complete skyrmion-like structure can be seen, associated with a negative region in the chirality maps. Other similar textures are incomplete due to system size and open boundary conditions.} 

 \end{itemize}
}

\subsection{{Quantum entanglement}}
\label{sec:entanglement}
 The preceding subsections involved expectation values of quantities that have a classical counterpart. We will now focus on quantum entanglement, which has no classical analogue. Over the last decades, entanglement has become a fundamental tool to characterise condensed matter physics systems with strong correlations \cite{JiangNature2012, Laflorencie2016, QIbook, SatooriNature2023, EntanglementWitnessReview2024}. Here, we aim to use this quantity to  inspect the quantumness of the obtained ground states. Since we are working with wavefunctions, which are pure states, we can use the entanglement entropy
$$ S_a=-\Trace(\rho_a \log_2\rho_a)$$
to study the entanglement between disjoint subsets $a$ and $b$ of the system ($a \cup b$), where $\rho_\text{a}=\Trace_b\rho$ is the reduced density matrix of subsystem $a$ obtained taking the partial trace over $b$ of the density matrix $\rho$ of the system. The density matrix $\rho$ associated with a ground state $\ket{\psi}$ is the projection operator given by $\rho=\ket{\psi}\!\bra{\psi}$.

\begin{figure}
    \centering
    \includegraphics[width=\linewidth]{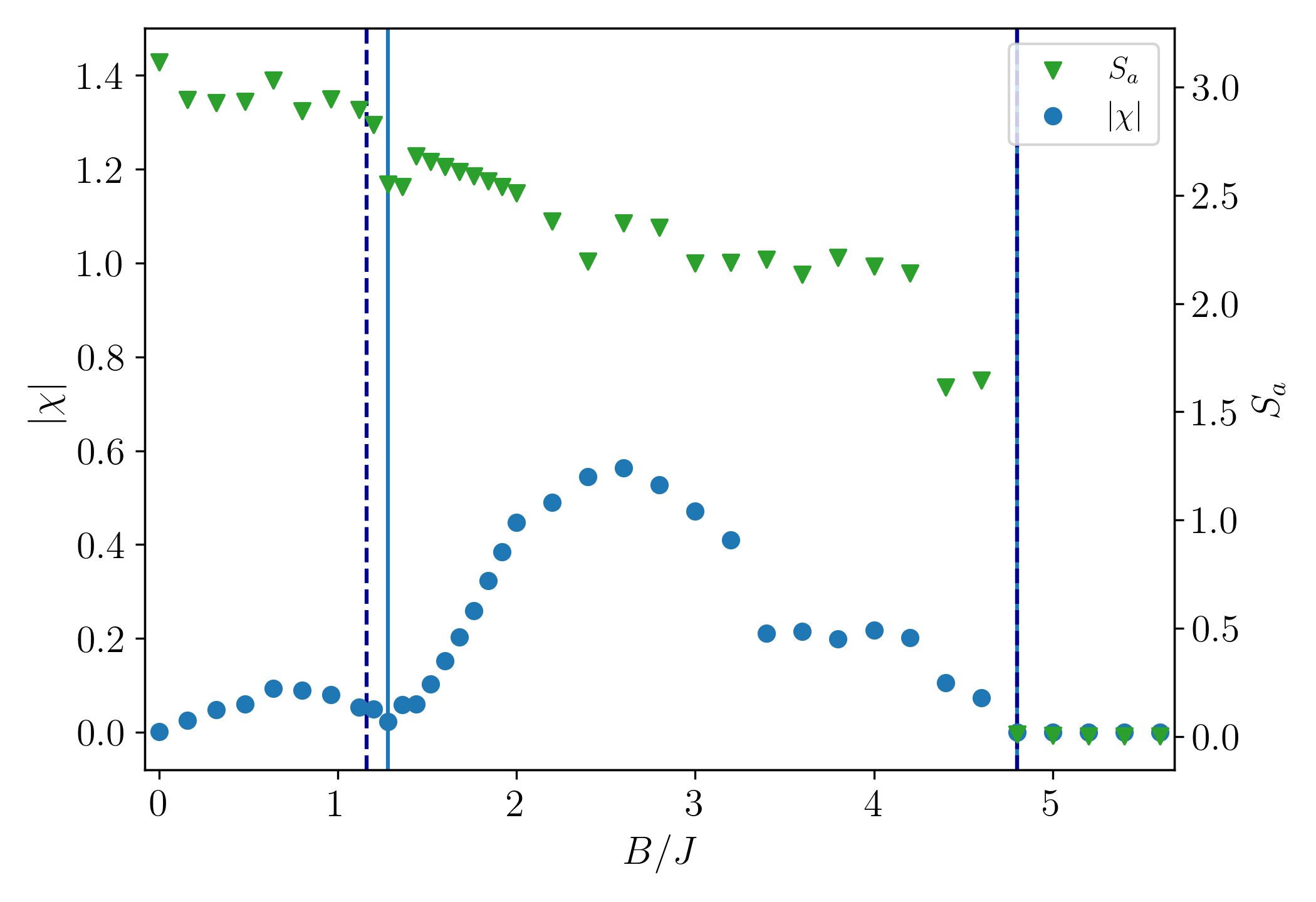}
    \caption{Entanglement entropy $S_a$ and rescaled chirality vs. $B/J$. Half-chain entanglement entropy (green triangles) was obtained considering each half of the system as subsystems A and B. Blue dots indicate {$|\chi|$} for each $B/J$. Both quantities vanish for $B/J\gtrsim 4.8$. Solid vertical lines indicate where {$\chi$} changes sign {at lower fields and goes to zero at higher fields (see Fig.~\ref{fig:Q_chi})} and dashed dark blue lines mark $B/J$ values in which the magnetic susceptibility is discontinuous.
    Both lines coincide at $B/J\simeq4.8$.} 
    \label{fig:Q_vs_S}
\end{figure}

To calculate $S_a$, we split our 1D-mapping of the 2D system (see Appendix \ref{ap:convergence}) into two contiguous subsets labeled as subsystems $a$ and $b$. In particular, we will take each equally-sized half of the system as subsystems $a$ and $b$. The entanglement entropy $S_a$ between each half of the system and for each value of $B/J$ is presented in Fig.~\ref{fig:Q_vs_S} along with a rescaling of {$\chi$}. The entanglement entropy is non-zero and almost constant at lower magnetic fields, drops to lower values at $B/J \simeq 1.2$, and becomes zero for $B/J \gtrsim 4.8$, which indicates that the ground state is a product state. This is consistent with the system being in a field-polarized phase at high magnetic fields. Moreover, $S_a$ takes non-overlapping values in each of the three regions delimited with solid vertical lines, which are the same three main regions delimited by the chirality (see Section \ref{sec:chirality}): $B/J \lesssim 1.2$, corresponding to AF helices, AF skyrmion and skyrmion-like textures in the $1.2 \lesssim B/J \lesssim 4.8$ range, and a field-polarized ground state for $B/J \gtrsim 4.8$. Additionally, we see that $S_a$ takes its highest values for the helical phase emerging at lower fields. This was also the case in the ferromagnetic system \cite{Qsky}.

  Based on these results, the entanglement entropy $S_a$ changes in the different types of ground states in this system.  Furthermore, we have shown it is clearly non-zero in the range of magnetic fields where topological textures are found, contributing to the quantum description of these ground states. In fact, in that region we also see there seem to be two subregions of $S_a$, in accordance to the two types of ground state textures at intermediate fields (Fig.~\ref{fig:Mz_variosB}(b) and (c)): AF skyrmions and a higher-field AF skyrmion-like texture, similar to that found in the classical system. If we take different left-right bipartitions of our system, we observe that $S_a \sim 10^{-3}$ in the fully-polarized state for each choice of $a$, whereas $S_a$ values are one order of magnitude larger for the AF skyrmion and helical phases.

\begin{figure}
\begin{minipage}{\linewidth}
    \centering
    \includegraphics[width=\linewidth]{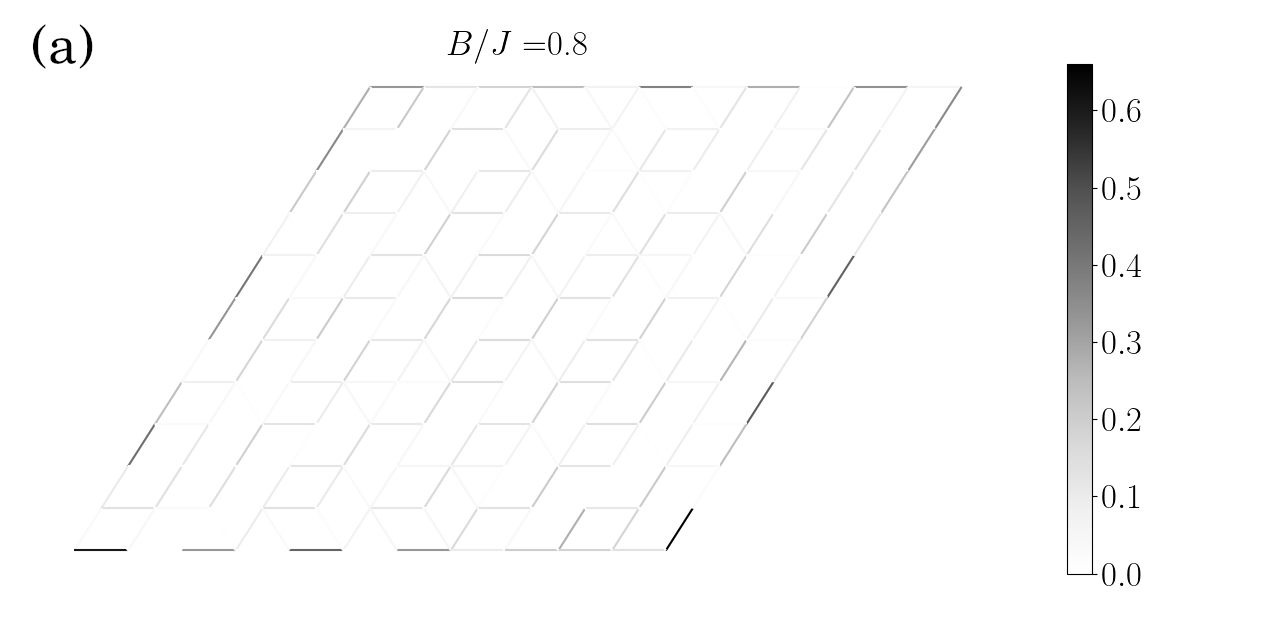}
\end{minipage}
\begin{minipage}{\linewidth}
    \centering
    \includegraphics[width = \linewidth]{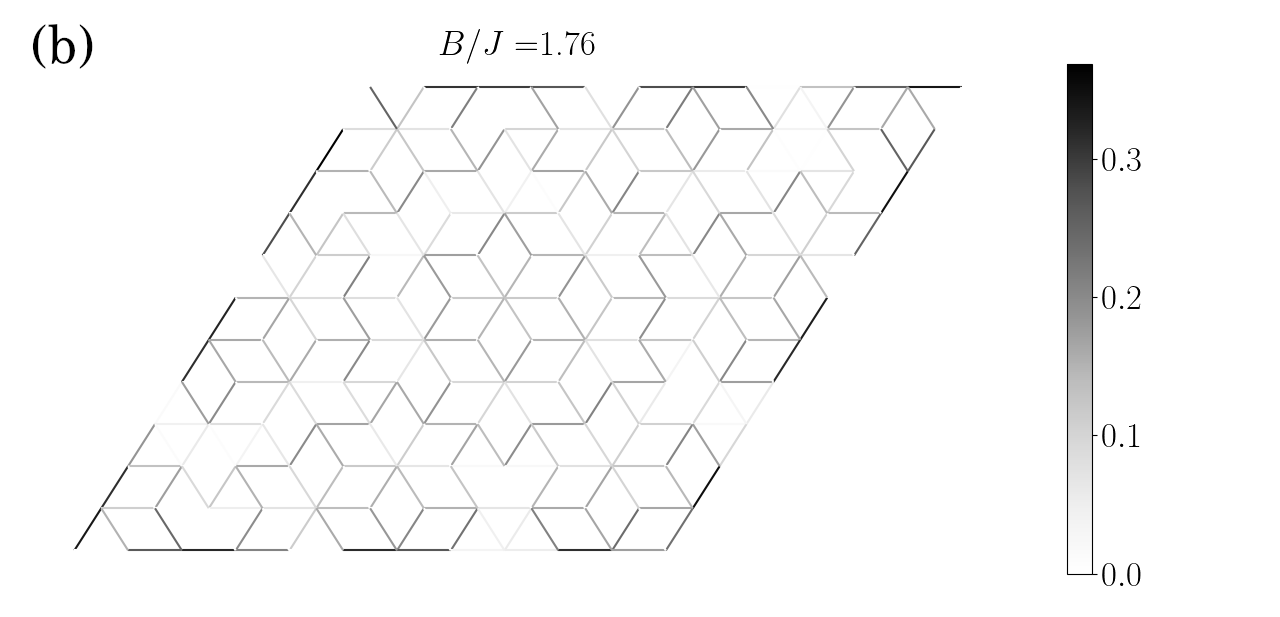}
\end{minipage}
\begin{minipage}{\linewidth}
    \centering
    \includegraphics[width = \linewidth]{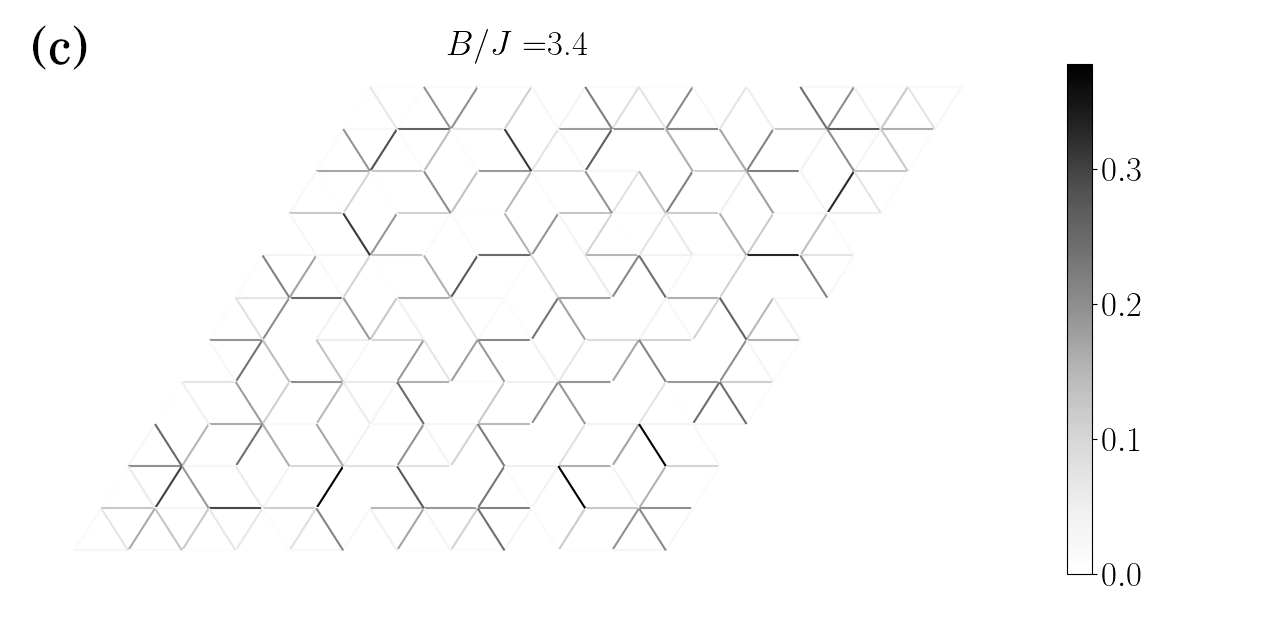}
\end{minipage}
\begin{minipage}{\linewidth}
    \centering
    \includegraphics[width = \linewidth]{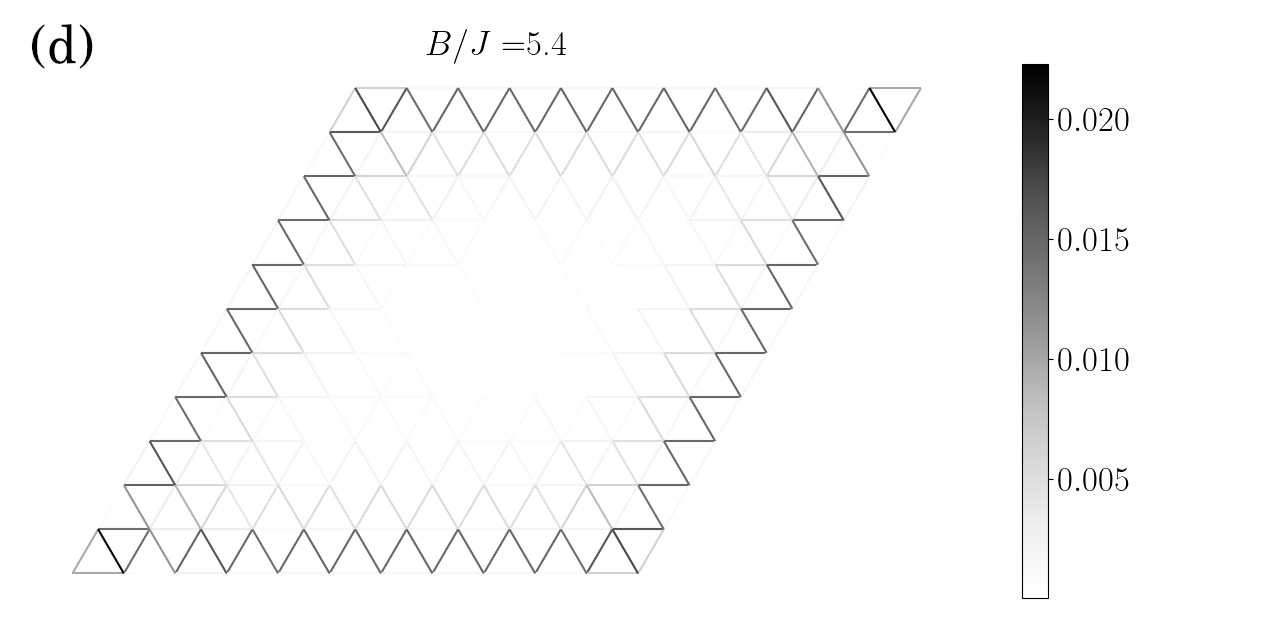}
\end{minipage}

\caption{Concurrence between nearest-neighbor spins for {(a)} the helical phase at $B/J=0.8$, {(b)} the AF skyrmion lattice at $B/J=1.76 $, {(c) the skymion-like phase at $B/J=3.4$} and {(d)} fully-polarized phase at $B/J=5.4$. Spin pairs involving at least one spin of the lattice boundary have the largest concurrences among all pairs in each phase considered. 
}

\label{fig:concurrence}
\end{figure}

So far we can conclude that there is entanglement in the AF skyrmion phase. In order to learn how this entanglement is distributed along the system, we employ the concurrence \cite{conc_wootters} between each pair of nearest-neighbor spins. To compute the concurrence $C_{ij}$ between spins $i$ and $j$, we first need the spin-flipped state  
$$\tilde{\rho}_{ij}=(\sigma_y \otimes \sigma_y){\rho}^*_{ij} (\sigma_y \otimes \sigma_y)$$ of the state $\rho_{ij}$, with $\sigma_y$ the Pauli matrix $y$ and ${\rho}^*_{ij}$ the complex conjugate of the reduced density matrix of spins $i$ and $j$. Then, $C_{ij}$ is given by
$$C_{ij}=\max\left(0, \sqrt{\lambda_1}-\sqrt{\lambda_2}-\sqrt{\lambda_3}-\sqrt{\lambda_4}\right), $$
where the $\lambda_i$'s are the eigenvalues of
$$R=\rho_{ij} \:\tilde{\rho}_{ij} $$ in decreasing order. Concurrence is such that $0\leq C_{ij} \leq 1$, where $C_{ij}=0$ for separable states and $C_{ij}=1$ if the pair is maximally entangled (e.g. if the pair is in a Bell state).

 Fig.~\ref{fig:concurrence} shows the concurrence between each pair of neighboring sites for different phases: AF helical, AF skyrmion, {skyrmion-like subphase (skyrmions without fully polarized cores)} and field-polarized phase. The AF helical texture displays the largest entanglement values of the {four} phases. {Remarkably, the highest concurrence values for the helical and ferromagnetic phases lie near the edges of the lattice, whereas in the skyrmion phases they are mostly distributed in the bulk.} Concurrences of neighboring pairs of the field-polarized state practically vanish in the bulk and reach values that are at most one order of magnitude smaller than the other phases.
 
Considering all of the above,  we can conclude that spins forming AF skyrmions are entangled, confirming the quantum nature of these ground states. Moreover,  judging by the entanglement entropy $S_a$, entanglement seems to change as a function of the magnetic field in the three main regions defined by  the magnetic susceptibility and the chirality. In particular,  our results of the concurrence $C_{ij}$ between spins $i$ and $j$ indicate that pair-wise entanglement notably differs in each of the main phases considered.

\subsection{ {Robustness of the antiferromagnetic skyrmion phase} 
}
\label{sec: stability}

\begin{figure*}[ht!]
    \includegraphics[width=0.8\linewidth]{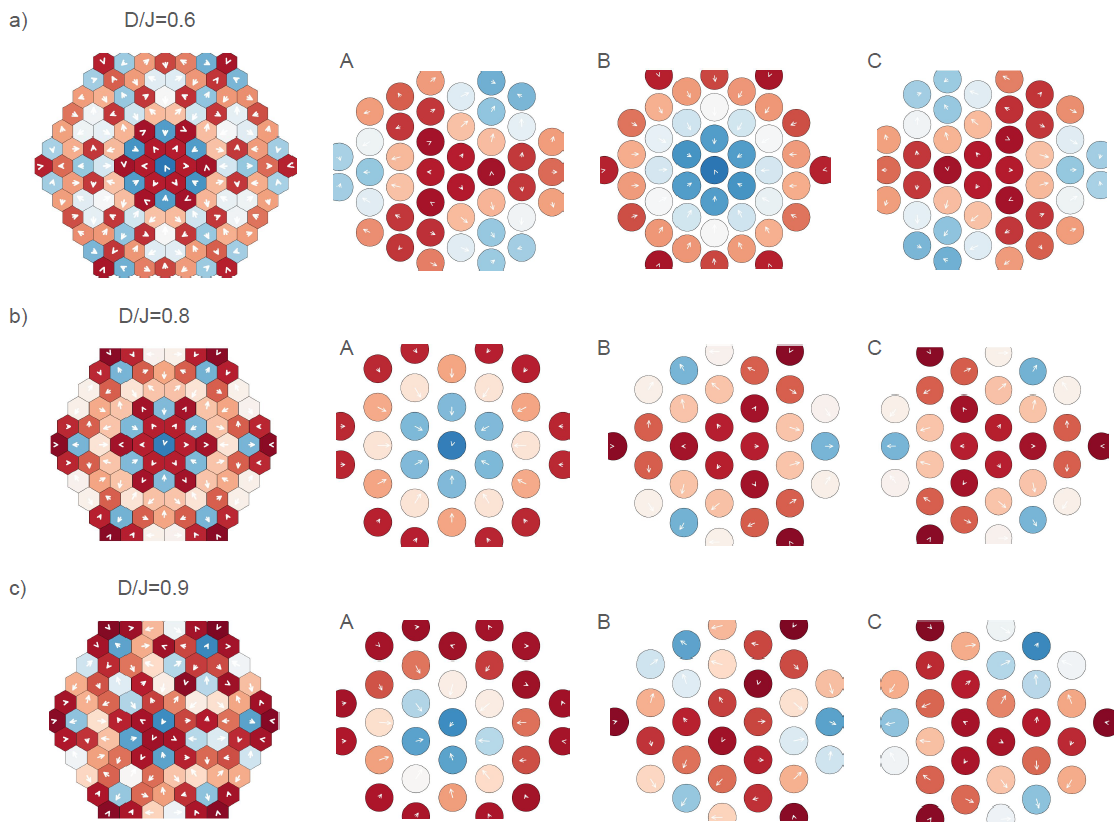}
    \centering
    
    \caption{Real-space magnetization per site of the ground states of Eq.\eqref{eq:H} (first column) and for each of the three sublattices (from second to fourth column) for different values of $D/J$ considering a disk-shaped triangular lattice. We take (a) $D/J=0.6$, $B/J=1.2$, (b) $D/J=0.8$, $B/J=1.76$, and (c) $D/J=0.9$, $B/J=2.0$. Notice that panel (b) matches the parameter set from Fig.~\ref{fig:Mz_variosB}(b).}

    \label{fig:distintosD}
\end{figure*}

\begin{figure*}[bt!]
    \centering
    \includegraphics[width=\linewidth]{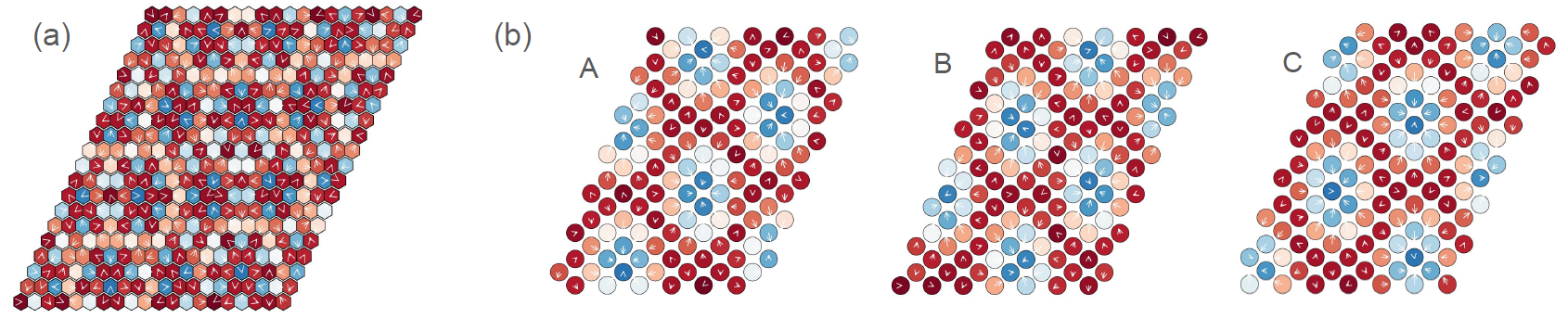}
    
    \caption{Local magnetization for the total lattice (panel a) and each sublattice (panel b) of the ground state of Eq.\eqref{eq:H} for $B/J=1.76$ and $D/J=0.8$ and considering a $L=20$ lattice.
    }
    \label{fig:subredesL20}
\end{figure*}

An important issue to address is the robustness  of the antiferromagnetic skyrmion phase with the shape and size of the lattice, given that boundary effects are expected due to the reduced system size. In order to study this, we redid the calculations for a triangular lattice with hexagon boundaries. In Fig.~\ref{fig:distintosD}(b) we show the resulting full lattice and sublattices textures for $B/J=1.76$  to compare them with Fig.~\ref{fig:Mz_variosB}(b). We can see that indeed there is a three-interpenetrated-sublattice antiferromagnetic skyrmion texture, and that the size of the sublattice skyrmions is roughly the same.

Moreover, to further explore the consistency of our calculations, we examine the local magnetization  for  other DMI values. These results are presented in Fig.~\ref{fig:distintosD}(a) and (c). It is well known that, in the classical case, skyrmion radii depend on the $D/J$ ratio: a larger (smaller) $D/J$ implies smaller (larger) skyrmions \cite{NagaosaTech}. This is also the case for these antiferromagnetic quantum skyrmions: larger skyrmions are found at $D/J=0.6$, $B/J=1.2$ (panel (a)) and smaller ones at $D/J=0.9$, $B/J=2.0$ (panel (c)).

{Additionally}, in Fig.~\ref{fig:subredesL20} we show that the antiferromagnetic skyrmion lattice is robust for larger system sizes, presenting the sublattice magnetization profiles for $D/J=0.8$, $B/J=1.76$ for $L=20$. Due to the larger system size, the three-sublattice structure is more noticeable. Furthermore, the size of the individual skyrmions in each sublattice does not change significantly.

{Finally, to check the robustness of the calculations  for different system sizes, we calculate what we define as ``bulk'' quantities of ground states at $B/J=1.76$ for different lattice sizes: the total quantum chirality, total classical chirality and the sum of the nearest-neighbor concurrence for a $9\times9$ patch of spins taken near the center of $L\times L$ lattices, with $L=12,15,18,20$. Results are presented in Table \ref{tab:chi}. In Appendix B the magnetization profiles and the chirality maps for the four different sizes used for these calculations are shown (see Fig.~\ref{fig:chipatches}), indicating the $9\times9$ section. 

Notably, the bulk values for larger system sizes do not vary significantly. This further supports the existence of a skyrmion phase for different system sizes. Bulk values for $L=12$ are smaller than for larger sizes, but this
is clearly due to boundary effects since a $9\times 9$ patch in a $12 \times 12$ lattice is very close to the borders. In fact, regions with large absolute value of $\chi$ near the lattice borders seem to fade away as $L$ increases. The dilution of these contributions with system size for the $9 \times 9$ patch can be appreciated in Fig.~\ref{fig:chipatches} in Appendix B.}

\begingroup

\setlength{\tabcolsep}{10pt}  
\renewcommand{\arraystretch}{1.5} 
\begin{table}[h!]
 {
    \centering
    \begin{tabular}{|c|c|c|c|}
       \hline
       $L$  & $\chi^{bulk}$ & $\chi^{bulk}_{clas}$ & $\sum C_{ij}^{bulk}$ \\
       \hline
       12  &  $ -0.14$ & $-0.24$ & $17.88$ \\
       \hline
       15  &  $ -0.28$ & $-0.32$ & $18.06$ \\
       \hline
       18  &  $-0.30$ & $-0.32$ & $18.22$ \\
       \hline
       20  & $-0.35$ & $-0.35$ & $18.6$\\
       \hline
    \end{tabular}
    
    \caption{Chirality and net concurrence of a patch of $9\times9$ near the center of lattices of $L\times L$ spins (patches are shown in Appendix B).}
    \label{tab:chi}
    }%
\end{table}
\endgroup

In summary, these analyses suggest that, despite some limitations due to the shape of the lattice, boundaries and system size,  our findings described in previous subsections {regarding antiferromagnetic skyrmions are} robust. It would be interesting to dive into a more systematic study of the different phases, particularly the skyrmion-like subphase, where there the textures in the sublattices are skyrmions without fully polarized cores. We defer this for future work. 

\section{Conclusions}
\label{sec:conclusions}

We have presented a study of a simple skyrmion-hosting antiferromagnetic model in the triangular lattice, combining exchange interactions and in-plane DMI under an external magnetic field. Through DMRG calculations, we have studied the ground states of this spin-1/2 model, focusing on the characterization and robustness of the three-sublattice antiferromagnetic skyrmion phase.

Our numerical findings imply the stabilization of AF skyrmions formed by three interpenetrated sublattices 
in a finite range of magnetic fields. Variables such as the local magnetization and chirality show the stabilization of these textures, which are similar to those found in its classical counterpart. {Local chirality maps support these results, showing clearly distinct patterns in the different phases.} Even though the spin textures at intermediate magnetic fields are reminiscent of classical AF skyrmions, we found that there is entanglement present in these states, which proves their quantum nature. Most interestingly, this quantity also serves as a parameter to distinguish this phase from the low-field helices and the high-field polarized states.

Moreover, we have checked the consistency of our calculations comparing results for different geometries of the (open) boundary conditions, and found that the AF skyrmion texture does not visibly change. We have seen that this is not a fine-tuned regime since these textures can be stabilized for different strengths of DMI as well, and the skyrmion size changes accordingly. Finally, although finite size effects are expected due to the computational limitations, we have seen that the AF skyrmion phase is present at larger systems. {In addition, we have checked that bulk calculations of relevant parameters such as the chirality and the concurrence are robust upon increasing the system size.}

  Given that one key feature of AF skyrmions is their topological protection, it would be interesting to explore how robust quantum AF skyrmions are against external perturbations. AF skyrmions also present no skyrmion Hall effect \cite{SkyAFHall1,SkyAFHall2,SkyFerri,Pham2024} under certain conditions \cite{Aldarawsheh2024}, which is of interest for technological applications, so a thorough study of the dynamics of quantum AF skyrmions would be pertinent. We leave both of these research avenues for future work.

\begin{acknowledgments}
I. C. would like to thank Mauricio Matera and Loic Herviou for insights and helpful discussions. The authors thank the Subsecretar\'ia de Ciencia y Tecnolog\'ia and the SNCAD from Argentina. I. C. also thanks the Laboratorio de Investigaci\'on y Formaci\'on en Inform\'atica Avanzada (UNLP-CICPBA, Argentina) for the computational resources. F. A. G. A. is partially supported by CONICET (PIP 2021-112200200101480CO, PIBAA 2872021010 0698CO), SECyT UNLP (PI+D X947 - X1065) and Agencia I+D+i (PICT-2020-SERIE A-03205). I. C., F. H. and L. R. are partially supported by SECyT UNLP (PI+D X943 - X1050). Part of the numerical simulations in this work were carried out using the Clementina XXI supercomputer from Argentina. 

\end{acknowledgments}

\appendix

\section{DMRG implementation and convergence}
\label{ap:convergence}
\counterwithin{figure}{section}
\setcounter{figure}{0}  

\begin{figure}[h!]
    \centering
    \begin{tikzpicture}
    \centering
    \foreach \row in {0, 1, ..., \rows} {
    \draw[very thick] ($\row*(\lattspacing*0.5, {\lattspacing*0.5*sqrt(3)})$) -- ($(\rows*\lattspacing,0)+\row*(\lattspacing*0.5, {\lattspacing*0.5*sqrt(3)})$);
    \draw[very thick] ($(\lattspacing*\row, 0)$) -- ($(\row*\lattspacing,0)+\rows*(0.5*\lattspacing, {0.5*\lattspacing*sqrt(3)})$);
    
    \draw[thick, red] ($({\lattspacing*\row - 0.05}, 0)$) -- ($({\row*\lattspacing - 0.05},0)+\rows*(0.5*\lattspacing, {0.5*\lattspacing*sqrt(3)})$); 
    }

    \draw[thick, red] ($(\lattspacing, 0)$) -- ($(\lattspacing, {0.5*\rows*\lattspacing*sqrt(3)})$);
    \draw[very thick] ($(0.5*\lattspacing, {0.5*\lattspacing*sqrt(3)})$) -- ($(\lattspacing, 0)$);
    \draw[very thick] ($2*(0.5*\lattspacing, {0.5*\lattspacing*sqrt(3)})$) -- ($(2*\lattspacing, 0)$);
    \draw[very thick] ($2*(0.5*\lattspacing, {0.5*\lattspacing*sqrt(3)}) + (\lattspacing,0)$) -- ($(2*\lattspacing,0)+ (0.5*\lattspacing,{0.5*\lattspacing*sqrt(3)})$);
    \draw[thick, red] ($({2*\lattspacing}, 0)$) -- ($({2*\lattspacing}, {0.5*\rows*\lattspacing*sqrt(3)})$);
    \filldraw[black] ({0.5*\lattspacing},{\lattspacing*0.5*sqrt(3)}) circle (3pt) node[anchor=north east, color=red]{2\:\:\:};
    \filldraw[black] (0,0) circle (3pt) node[anchor=north east, color=red]{1\:\:};
    \filldraw[black] (\lattspacing,0) circle (3pt) node[anchor=north east, color=red]{4\:\:\:};
    \filldraw[black] ({1.5*\lattspacing},{\lattspacing*0.5*sqrt(3)}) circle (3pt) node[anchor=north east, color=red]{5\:\:};
    \filldraw[black] (\lattspacing,{\lattspacing*sqrt(3)}) circle (3pt) node[anchor=north east, color=red]{3\:\:\:};
    \filldraw[black] ({2*\lattspacing},0) circle (3pt) node[anchor=north east, color=red]{7\:\:\:};
    \filldraw[black] ({2*\lattspacing},{\lattspacing*sqrt(3)}) circle (3pt) node[anchor=north east, color=red]{6\:\:\:};
    \filldraw[black] ({2.5*\lattspacing},{\lattspacing*0.5*sqrt(3)}) circle (3pt) node[anchor=north east, color=red]{8\:\:\:};
    \filldraw[black] ({3*\lattspacing},{\lattspacing*sqrt(3)}) circle (3pt) node[anchor=north east, color=red]{9\:\:};
    
    \end{tikzpicture}
    \caption{Triangular lattice (black lines) and its associated 1D chain (red lines).}
    \label{fig:1Dmap}
\end{figure}
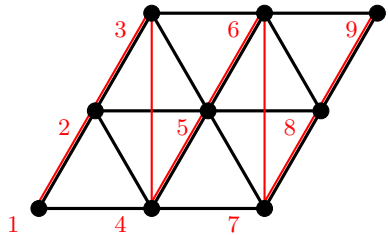

DMRG was introduced as a numerical approach  to study one-dimensional quantum systems. Therefore, we must map each 2D lattice into a 1D chain to apply the MPS formulation of this algorithm. The goal is to choose a mapping in which a fair amount of the neighboring spins from the lattice are still nearest neighbors in the chain since the 2D-to-1D mapping inherently introduces long-ranged interactions in the resulting 1D model. In our case, we use a zigzag ordering as in Ref.~\cite{Qsky}. An example of a triangular lattice mapped into a 1D chain using this ordering is presented in Fig.~\ref{fig:1Dmap}. 

\begin{figure*}[th!]
    \begin{minipage}{0.32\linewidth}
    \includegraphics[width=\linewidth]{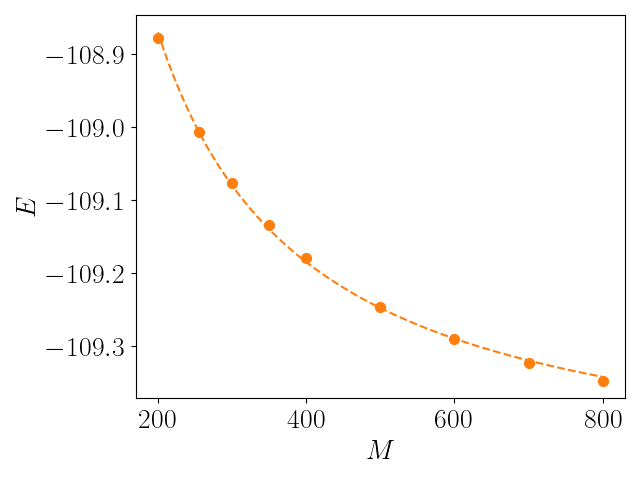}
    \end{minipage}
    \hfill
    \begin{minipage}{0.32\linewidth}
    \includegraphics[width=\linewidth]{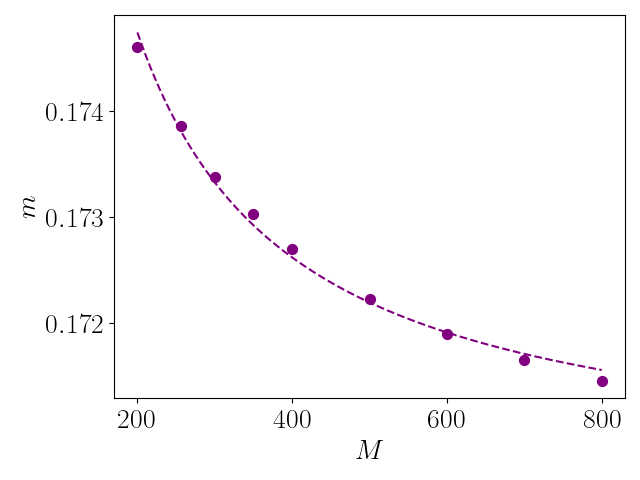}
    \end{minipage}
    \hfill
    \begin{minipage}{0.32\linewidth}
    \includegraphics[width=\linewidth]{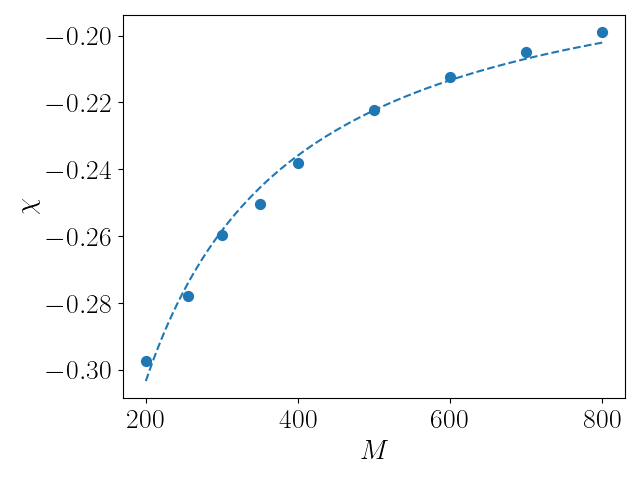}
    \end{minipage}
    \caption{Energy (left), average magnetization per site (center) and chirality (right) vs bond dimension {$M$} of ground states of Eq.\eqref{eq:H} for $B/J=1.76$. The system hosts an AF skyrmion phase at this value of $B/J$ as shown in Fig.~\ref{magn}. 
    }
    \label{fig:BDstudy}
\end{figure*}

In the DMRG calculations, we used a random MPS as initial state and performed a large number of sweeps at smaller bond dimensions {$M$ ($M\leq100$)} to aid convergence. Then, we employed gradually larger values of {$M$} until reaching the prestablished maximum value. From that point onwards, sweeps continued until the relative energy difference between sweeps was smaller than $10^{-7}$. Taking into account that DMRG is a variational algorithm, we repeated several times the ground state simulations for each $B/J$ and different values of {$M$}. Our results of the main text correspond to ground state approximations $\psi(B)$ in which $E$ and $m$ change in less than $0.2\%$ between subsequent {$M$}'s.

Since we are especially interested in the AF skyrmion phase, we check for convergence of several properties with respect to the bond dimension {$M$} for $B/J=1.76$, where the ground state is a three-sublattice AF skyrmion texture. Due to the size of the system, we expect there to be boundary effects, especially in the chirality calculations, so in order to study the bulk chirality we did not include boundary sites in the computation of {$\chi$}. In Fig.~\ref{fig:BDstudy} we plot the energy, magnetization and chirality of the ground state obtained for a $L=12$ system at $B/J=1.76$ as a function of the bond dimension {$M$}. In all three cases, we see that clearly the system converges to a finite value. Fitting the data with a simple hyperbolic function, we find that these quantities tend to $E_{{M} \to \infty}=-109.50024(2)$, $m_{{M} \to \infty}=0.170500080(5)$ and ${\chi_{M \to \infty}}=-0.168346(9)$, respectively. 

\section{{Patches used for the calculation of ``bulk'' quantities and chirality maps for different system sizes}}
\label{ap:chirality}
\begin{figure*}

\centering
    \includegraphics[width = 0.8\textwidth]{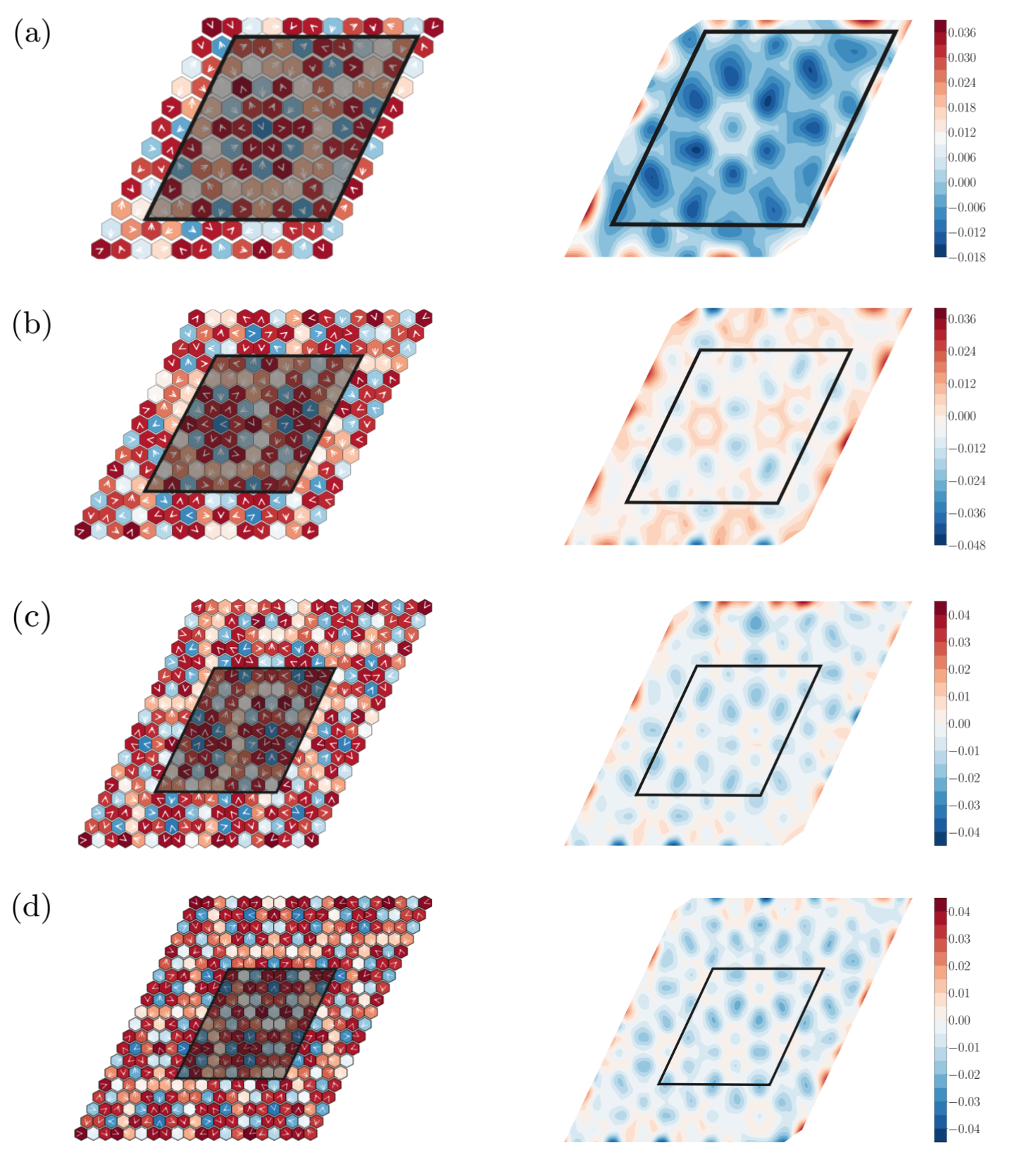}
\caption{{Patches used to compute $\chi^{bulk}$ and $\chi_{clas}^{bulk}$ in Table \ref{tab:chi} (left column) and quantum local chirality (right column) for each $L\times L$ lattice, with (a) $L=12$, (b) $L=15$, (c) $L=18$ and (d) $L=20$. As we can see, there are less extreme values of $\chi(i)$ near the edges as $L$ increases. Notice that each map has its own colorbar.}}
\label{fig:chipatches}
\end{figure*}

{In Section \ref{sec: stability} of the main text, we compute the chirality and the concurrence for $9\times 9$ patches for the skyrmion-like phases in different system sizes to show the robustness of the phase. The patches of $9\times 9$ spins chosen for each lattice size $L$ are presented in Fig.~\ref{fig:chipatches}.

The use of bulk quantities computed in these regions of $9\times 9$ spins allows us to make a quantitive comparison of the behavior of the skyrmion phase for different system sizes. Nevertheless, a qualitative comparison is also possible through the analysis of chirality maps for each $L$. In Fig.~\ref{fig:chilocal} we show the local quantum chirality of the AF skyrmion lattice for $L=18$. Here we include for completeness all the quantum chirality maps for the system sizes considered in Table \ref{tab:chi}.}

\providecommand{\noopsort}[1]{}\providecommand{\singleletter}[1]{#1}%

\end{document}